\documentclass[aps,onecolumn,preprint,superscriptaddress
]{revtex4-1}
\pdfoutput=1
\usepackage{amsmath}
\usepackage{amssymb}
\usepackage{subfig}
\usepackage{graphicx}
\usepackage{float}
\usepackage{tabularx}
\usepackage{hyperref}
\graphicspath{{"./arxiv_figures/"}}
\begin{document}

%\preprint{}

\title{Unifying first principle theoretical predictions and experimental measurements of size effects on thermal transport in SiGe alloys}


\author{Samuel Huberman}
%\thanks{}
\affiliation{Department of Mechanical Engineering, Massachusetts Institute of Technology, Cambridge, Massachusetts 02139, USA}
\author{Vazrik Chiloyan}
\affiliation{Department of Mechanical Engineering, Massachusetts Institute of Technology, Cambridge, Massachusetts 02139, USA}
\author{Ryan A. Duncan}
\affiliation{Department of Chemistry, Massachusetts Institute of Technology, Cambridge, Massachusetts 02139, USA}
\author{Lingping Zeng}
\affiliation{Department of Mechanical Engineering, Massachusetts Institute of Technology, Cambridge, Massachusetts 02139, USA}
\author{Roger Jia}
\affiliation{Department of Materials Science and Engineering, Massachusetts Institute of Technology, Cambridge, Massachusetts 02139, USA}
\author{Alexei A. Maznev}
\affiliation{Department of Chemistry, Massachusetts Institute of Technology, Cambridge, Massachusetts 02139, USA}
\author{Eugene A. Fitzgerald}
\affiliation{Department of Materials Science and Engineering, Massachusetts Institute of Technology, Cambridge, Massachusetts 02139, USA}
\author{Keith A. Nelson}
\affiliation{Department of Chemistry, Massachusetts Institute of Technology, Cambridge, Massachusetts 02139, USA}

\author{Gang Chen}
\email[]{gchen2@mit.edu}
\affiliation{Department of Mechanical Engineering, Massachusetts Institute of Technology, Cambridge, Massachusetts 02139, USA}


\date{\today}

\begin{abstract}
In this work, we demonstrate the correspondence between first principle calculations and experimental measurements of size effects on thermal transport in SiGe alloys. Transient thermal grating (TTG) is used to measure the effective thermal conductivity. The virtual crystal approximation under the density functional theory (DFT) framework combined with impurity scattering is used to determine the phonon properties for the exact alloy composition of the measured samples. With these properties, classical size effects are calculated for the experimental geometry of reflection mode TTG using the recently-developed variational solution to the phonon Boltzmann transport equation (BTE), which is verified against established Monte Carlo simulations. We find agreement between theoretical predictions  and experimental measurements in the reduction of thermal conductivity (as much as $\sim$ 25\% of the bulk value) across grating periods spanning one order of magnitude. This work provides a framework for the tabletop study of size effects on thermal transport.
\end{abstract}

%\keywords{}

\maketitle

\section{Introduction}\label{sec:intro}
Silicon-Germanium (SiGe) alloys are the canonical example for the study of thermal transport in a mass-disordered, yet crystalline system as evidenced by the plethora of work, dating back to the original work by Stohr~\cite{stohr1939two} and Abeles~\cite{abeles1962thermal,abeles1963lattice}, where it was noted that the the mass-disorder scatters short-wavelength phonons consequently shifting the dominant contribution to thermal conductivity to long wavelength phonons.

On the theoretical side, earlier works relied on empirical models. Skye and Schelling used molecular dynamics to study the relative contribution between mass and bond disorder, finding larger resistivity than experiment \cite{skye2008thermal}. Bera et al. estimated the mean free paths (MFP) of phonons in SiGe using a model based on the expected scalings of the phonon lifetimes~\cite{bera2010marked}. Significant progress took place when Garg et al. demonstrated the viability of first principle approaches to estimate the bulk thermal conductivity of SiGe~\cite{garg2011role}. Recently, Iskandar modeled thin SiGe films to include the effect of boundary modes~\cite{iskandar2015interplay}.

On the experimental side, Koh et al. reported a modulation frequency dependent estimate of thermal conductivity under the Fourier model of the experimental geometry of time domain thermoreflectance (TDTR)~\cite{koh2007frequency}. The authors proceeded to argue that the frequency dependence corresponds to suppression of phonons with MFP greater than the thermal penetration depth. This result led to a series of theoretical explanations. da Cruz et al. presented a framework to divide thermal transport into harmonic and anharmonic channels~\cite{da2012role}. Vermeersch et al. explained the divergence from bulk using truncated Levy walks~\cite{vermeersch2015superdiffusive, vermeersch2015superdiffusive2}. Recently, Hua et al. have found that a microscopic model of the interface between the aluminum transducer and the substrate can be used to explain away this frequency dependence and recover a bulk value of SiGe~\cite{hua2016experimental}. Wilson et al. reported decreasing thermal conductivity and increasing interface conductance with increasing modulation frequency and argue that the interplay between the contributions of long wavevector phonons and interfacial scattering is responsible for this observation~\cite{wilson2014anisotropic}.

The objective of this work is to provide a framework upon which size effects on thermal transport can be probed and understood with tabletop experiments. This is accomplished using a bottom-up theoretical approach combined with a simple experimental geometry that is not obfuscated by the interface present in TDTR; the transient thermal grating (TTG) technique. In doing so, we are able to unify the pictures obtained from the macroscopic observables of experiment to the microscopic properties from theory.

The structure of the paper is as follows. In Section~\ref{sec:theory} we present the phonon properties obtained using density functional theory. In Section~\ref{sec:BTE}, the variational solution to the phonon BTE for the TTG experimental geometry is developed. In Section ~\ref{sec:experiment}, results obtained from TTG are presented and compared with our BTE-based predictions. Finally, we close with a look towards future work in Section~\ref{sec:discussion}.

\section{\label{sec:theory} Theory}
\subsection{First Principle Calculations}
We follow the general procedure established by Broido~\cite{broido2005lattice,broido2007intrinsic} and Esfarjani~\cite{esfarjani2011heat}, to obtain the phonon properties for SiGe. While the details can be found in these works, an outline of the procedure is included for the sake of completeness.

For a non-alloy system, the harmonic phonon properties (dispersion, heat capacity, group velocity) are obtained using density-functional perturbation theory (DFPT). The underlying premise is to treat the mechanical displacement corresponding the wavevector of a phonon as a linear perturbation to the electronic Hamiltonian, from which atomic forces can be calculated under the self-consistent criteria of DFT. These forces are then converted into harmonic force constants and used to construct the dynamical matrix for the perturbing wavevector, which can then be diagionalized to obtain the corresponding frequencies.

The anharmonic properties, namely lifetimes (but also frequency shifts) can be obtained by extending the perturbation to higher orders. An alternative approach is to construct a symmetry-reduced set (based on the space group of the lattice) of atomic displacements in a supercell, where each member of the set undergoes a standard DFT self-consistent calculation, each yielding the force field for the configuration. With this set of force fields, the third order force constants are extracted (as solutions to a set of linear equations). Phonon lifetimes are related to the third order force constants through the application of Fermi's golden rule.

Integrating the modal thermal conductivity over the Brouillin zone, under the relaxation time approximation to the phonon Boltzmann transport equation (BTE), yields the lattice thermal conductivity. This full procedure is made concrete with the ShengBTE package~\cite{li2014shengbte}.

To extend the above procedure to a crystalline alloy, approximations are necessary. Following Garg et al.~\cite{garg2011role}, we use the virtual crystal approximation (VCA). Within this approximation, two paths can be taken. One can compositionally average the pseudopotentials for the constituent atoms, and then proceed with the usual procedure. Alternatively, one can calculate the harmonic and third order force constants for the unalloyed crystalline versions of the constituent atoms, take the mass normalized compositional average and then proceed to calculate the phonon properties:

\begin{equation}
A_{VCA} = x A^{Si} + (1-x) A^{Ge}
\end{equation}
where $A$ is a placeholder for the harmonic force constants, the third order force constants, the atomic masses and the lattice constants~\cite{antonius2016temperature}. We have followed both VCA procedures, and find negligible difference in the phonon properties (see supplementary material).

The penultimate step in the alloy calculation is to include the effect of mass disorder. Again, following Garg's work, the phonon lifetimes are modified under Matthiessen's rule using the theory established by Tamura~\cite{tamura1983isotope} to treat isotope scattering as an elastic perturbation through the coupling parameter $g$ defined as:

\begin{equation}
g = \sum_{i \in Si,Ge}f_i (1 - \frac{m_i}{m_{vca}})^2
\end{equation}
where $f_i$ is the concentration and the scattering time scales as $\tau_{disorder} ~\sim \frac{1}{g}$. Garg et al. went a step further to estimate the anharmonic shifts do due disorder through supercell calculations. Feng et al. used molecular dynamics to show that the application of Matthiessen's rule leads to an overestimation of thermal conductivity in SiGe due to neglecting four and five-phonon processes~\cite{feng2015coupling}. Our experimental results will show that the harmonic mass disorder approximation under Matthiessen's rule is a sufficient approximation. We note that this procedure will not capture the frequency shifts that can be observed in the SiGe Raman spectra~\cite{feldman1966raman,sui1993raman} (see the supplementary material). It is expected that these modes do not significantly contribute to thermal conductivity, as their group velocities are small and their lifetimes have been reduced by mass disorder scattering. The virtual crystal approximation is expected to break down when the mass disorder takes on a correlation length on the order of the probing length scale~\cite{mendoza2015ab}.

The DFT calculation parameters used in this work are the following: for the DFPT portion, a 16 $\times $ 16 $\times$ 16 Monkhorst-Pack $k$ mesh with a kinetic energy cutoff of 50 Ry and a convergence criteria of 1E-12 Ry is used. For the supercell calculations, a 4 $\times $ 4 $\times$ 4 supercell was used such that third order force constants up to the fifth nearest neighbor could be obtained and only wavefunctions at the gamma point were calculated. Both (Si,Ge).pz-bhs.UPF and (Si,Ge).pz-n-nc.UPF pseudopotentials were tested yielding a negligible difference between thermal conductivity estimates (see supplementary material). The DFPT calculations were done with a 6 $\times $ 6 $\times$ 6 $q$ mesh. Interpolation was done on a 48 $\times $ 48 $\times$ 48 $q$ mesh with a Gaussian smearing parameter of 0.1 for the Kronecker delta approximation to yield convergence of the thermal conductivity. All calculations were done with the quantum-ESPRESSO package~\cite{giannozzi2009quantum}. The input files and the properties are available in the supplementary material.

\begin{figure}
\centering     %%% not \center
\includegraphics[width=0.66\textwidth]{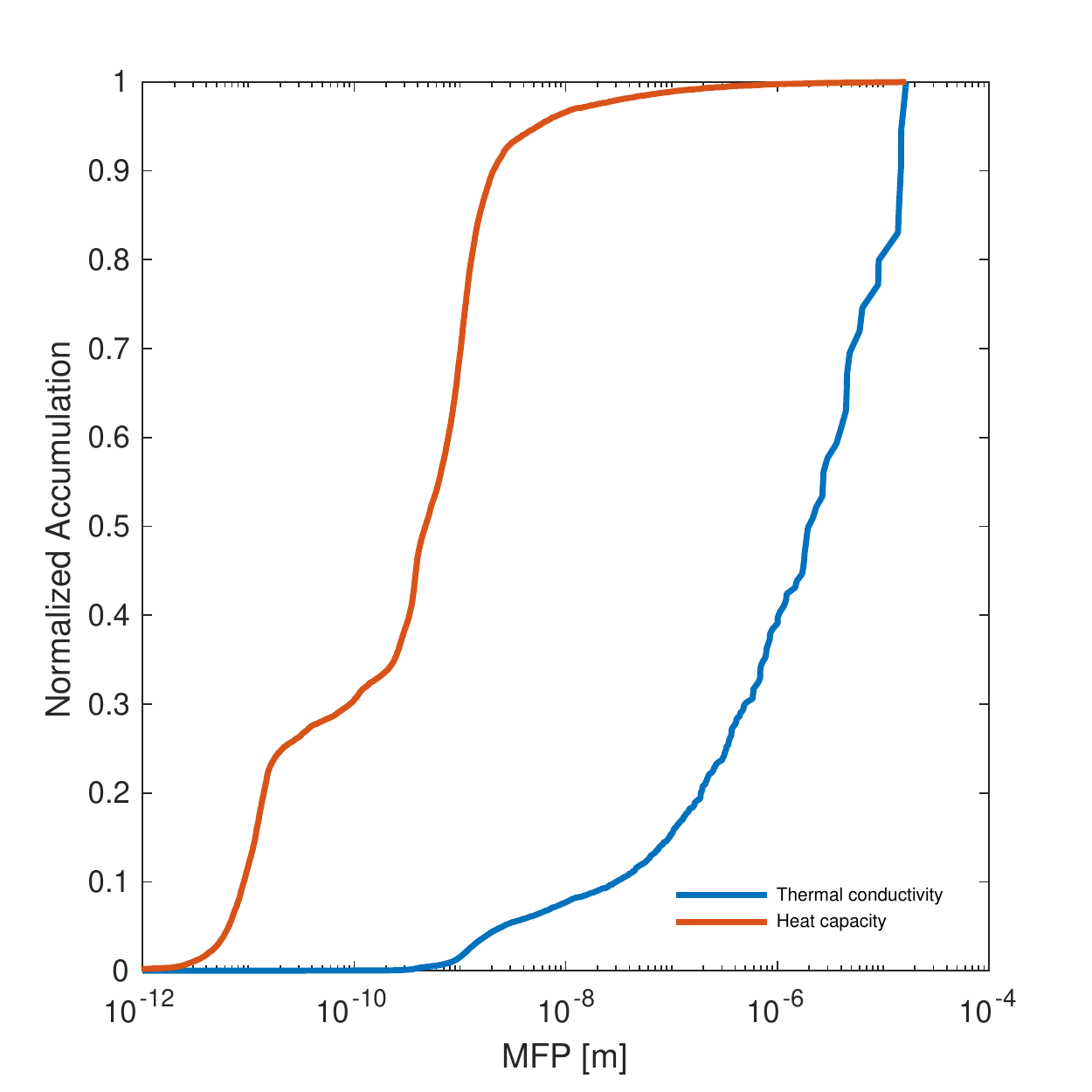}
\caption{Si$_{93.4}$Ge$_{6.6}$ MFP accumulation at 300 K.}
\end{figure}

\subsection{Solving the Boltzmann Transport Equation}\label{sec:BTE}
Given the bulk phonon properties of Si$_{93.4}$Ge$_{6.6}$, we now turn to the study of classical size effects on thermal transport in the reflection mode TTG geometry. The diffusive temperature profile has been solved previously in order to analyze the temperature signal using TTG for opaque materials~\cite{johnson2012phase}. For the experimental conditions of a spatially periodic heat source defined by wavevector $q = \frac{2\pi}{\lambda}$, the temperature is given by $T(x,z,t) = T_0 + T_0e^{iqx}h(z,t)$ in complex form, and this serves as a definition of the non-dimensional temperature $h$. The temperature $T_0$ is the background equilibrium temperature of the system, for example the room temperature. The heating by the laser is incorporated with a volumetric heat generation term, given by the functional form: 

\begin{equation}
Q = \delta(t) \mathrm{e}^{iqx}U_0 \beta \mathrm{e}^{-\beta z}
\end{equation}
where $U_0$ represents the energy per unit area deposited into the substrate by the pulse, and $\beta$ is the inverse penetration depth of the heating profile. The derivation found in~\cite{johnson2012phase} takes into consideration different in-plane and cross-plane thermal conductivities, however the experimental signal is sensitive to the in-plane thermal conductivity. For simplicity, we show the derivation for an isotropic system, where the Fourier heat conduction equation simplifies to:

\begin{equation}
\frac{\partial h}{\partial t} = -\alpha q^2 h + \alpha \frac{\partial^2 h}{\partial z^2} + \frac{\beta U_0}{CT_0}\mathrm{e}^{-\beta z}\delta(t)
\end{equation}
with the initial and boundary conditions given by: 

\begin{equation}
\begin{split}
h(z,t = 0^-) & = 0\\
\frac{\partial h}{\partial z}|_{z=0} & = 0 \\
h(z \to \infty, t) &= 0 
\end{split}
\end{equation}
which assumes an adiabatic surface at $z = 0$, and that the system starts at equilibrium prior to the energy deposited by the laser. We present the solution in the Laplace transformed domain for convenience:

\begin{equation}\label{eq:fourierTemp}
\hat{h}(z,s)= \frac{\frac{\beta U_0}{CT_0}}{s + \alpha(q^2-\beta^2)}\left(\mathrm{e}^{-\beta z} - \frac{\beta}{\sqrt{q^2+\frac{s}{a}}}\mathrm{e}^{-z\sqrt{q^2+\frac{s}{a}}}\right)
\end{equation} 
 
We intend to utilize this Fourier heat conduction temperature profile in our variational solution of the BTE. Taking the inverse Laplace transform of this yields the temperature as a function of the depth into the substrate and time: 

\begin{equation}\label{eq:fourierTemp2}
h(z,t) = \frac{\beta U_0}{2CT_0} \mathrm{e}^{-\alpha t(q^2-\beta^2)} \left(\mathrm{e}^{\beta z} \mathrm{erfc}(\beta\sqrt{\alpha t} + \frac{z}{2\sqrt{\alpha t}}) + \mathrm{e}^{-\beta z} \mathrm{erfc}(\beta\sqrt{\alpha t} - \frac{z}{2\sqrt{\alpha t}})\right)
\end{equation} 
where the surface heating profile is: 
 
\begin{equation}\label{eq:fourierTempSurf}
h(z=0,t) = \frac{\beta U_0}{CT_0} \mathrm{e}^{-\alpha t(q^2-\beta^2)} \mathrm{erfc}(\beta\sqrt{\alpha t})
\end{equation} 

\subsubsection{Temperature integral equation}
We begin with the spectral Boltzmann transport equation under the relaxation time approximation (RTA):

\begin{equation}
\frac{\partial g_{\omega}}{\partial t} + \mathbf{v_{\omega}}\cdot \mathbf{\nabla}g_{\omega} = \frac{g_0-g_{\omega}}{\tau_{\omega}} + \frac{Q_{\omega}}{4\pi}
\end{equation}
where $g_{\omega}$ is the phonon energy density per unit frequency interval per unit solid angle above the reference background energy, related to the distribution function as $g_{\omega} = \frac{\hbar \omega D(\omega)}{4 \pi} (f_{\omega}-f_0(T_0))$. $\mathbf{v_{\omega}}$ is the group velocity, $\tau_{\omega}$ is the relaxation time, and $g_0$ is the equilibrium energy density, given by $g_0 \approx \frac{1}{4\pi} C_{\omega}(T-T_0)$ in the linear response regime. The sinusoidal heating profile in the $x$-direction (in-plane), given by the pulse form $Q_{\omega}(x,z,t) = \delta(t)\mathrm{e}^{iqx}\tilde{Q_{\omega}}(z)$, means we can expect that the spectral and equilibrium energy densities to also obey a sinusoidal profile $g_{\omega}=\mathrm{e}^{iqx}\tilde{g_{\omega}}$ and the equilibrium distribution will simplify accordingly to $\tilde{g_{0}} = \frac{C_{\omega}T_0}{4 \pi}h(z,t)$. By inputting this in-plane sinusoidal profile and utilizing the Laplace transform (denoted by the $\hat{}$ symbol) in the time domain, the BTE simplifies to: 

\begin{equation}
\frac{\partial \hat{\tilde{g}}_{\omega}}{\partial z} + \hat{\tilde{g}}_{\omega} \frac{1 + s\tau_{\omega}+i\eta_{\omega}\mu_x}{\Lambda_{\omega}\mu_z}= \frac{\hat{\tilde{g}}_{0}+ \tau_{\omega}\frac{\tilde{Q_{\omega}}}{4\pi} }{\Lambda_{\omega}\mu_z}
\end{equation}
where we have defined $\eta_{\omega} = q\Lambda_{\omega}$. For convenience, we define the parameter $ V = \frac{1 + s\tau_{\omega}+i\eta_{\omega}\mu_x}{\Lambda_{\omega}\mu_z} $ to group the variables in a compact form for the following solution of the BTE:

\begin{equation}
\hat{\tilde{g}}_{\omega}(z,s,\mu_x,\mu_z) = \mathrm{e}^{-Vz}\hat{\tilde{g}}_{\omega}(z = 0,s,\mu_x,\mu_z) + \int_0^z dz' \mathrm{e}^{-V(z-z')}\frac{\hat{\tilde{g}}_{0}(z',s)+ \tau_{\omega}\frac{\tilde{Q_{\omega}}}{4\pi} }{\Lambda_{\omega}\mu_z}
\end{equation}
The boundary conditions are taken to be:

\begin{equation}
\begin{split}
\hat{\tilde{g}}_{\omega}(z=L,s,\mu_x,\mu_z<0) &= 0\\
\hat{\tilde{g}}_{\omega}(z=0,s,\mu_x,\mu_z>0) &= \sigma
\end{split}
\end{equation}
 
The first boundary condition takes an imaginary blackbody wall at length $L$ into the substrate at the background temperature to account for the semi-infinite substrate, where this length will limit to infinity. The second boundary condition provides the adiabatic boundary condition with diffuse scattering, where $ \sigma = \frac{1}{\pi}\int d\Omega \Theta (\mu_z) \mu_z \hat{\tilde{g}}_{\omega}(z=0,s,\mu_x,-\mu_z)$, which is proportional to the specular heat flux approaching the surface. We have utilized the Heaviside step function to reduce the integration over the solid angle only to consider phonons approaching the surface. Solving the boundary conditions, and taking the artificial length $L$ to infinity yields the formal solution to the BTE for the spectral energy density in terms of the equilibrium energy density:

\begin{equation}\label{eq:spectralBTE}
\begin{split}
\hat{\tilde{g}}_{\omega}(z,s,\mu_x,\mu_z) &= -\Theta(-\mu_z) \int_z^\infty \mathrm{e}^{-V(z-z')}\frac{\hat{\tilde{g}}_{0}(z',s)+ \tau_{\omega}\frac{\tilde{Q_{\omega}}}{4\pi} }{\Lambda_{\omega}\mu_z} \\
&+ \Theta(\mu_z) \left(\int_0^z \mathrm{e}^{-V(z-z')}\frac{\hat{\tilde{g}}_{0}(z',s)+ \tau_{\omega}\frac{\tilde{Q_{\omega}}}{4\pi} }{\Lambda_{\omega}\mu_z} + \int_0^\infty 2\mathrm{e}^{-V(z)}F_2(z')\frac{\hat{\tilde{g}}_{0}(z',s)+ \tau_{\omega}\frac{\tilde{Q_{\omega}}}{4\pi} }{\Lambda_{\omega}}\right)
\end{split}
\end{equation} 
where we have defined the following solid angle integral function: 

\begin{equation}
F_n(z) = \frac{1}{2 \pi}\int d\Omega \Theta(\mu_z)\mu_z^{n-2}\mathrm(e)^{-Vz}
\end{equation} 
The first term represents phonons moving towards the surface of heating at $z = 0$, whereas the second term represents phonons moving away from the surface.

The temperature equation can be derived by utilizing the equilibrium condition obtained by integrating Eq.~\ref{eq:spectralBTE} with respect to frequency and the solid angle~\cite{majumdar1993microscale}. The equilibrium condition in this case can be expressed as: 

\begin{equation} \label{eq:equilBTE}
4 \pi \int d \omega \frac{1}{\tau_\omega}\hat{\tilde{g}}_{0}(z,s) = \int d \omega \frac{1}{\tau_\omega} \int d \Omega \hat{\tilde{g}}_{\omega}(z,s,\mu_x,\mu_z)
\end{equation} 

Performing the solid angle integral, and inputting the expression for the non-dimensional temperature expression $\hat{\tilde{g_{0}}} = \frac{C_{\omega}T_0}{4 \pi}\hat{h}(z,s) $, we obtain the integral equation for the temperature distribution:

\begin{equation}\label{eq:intBTE}
\hat{h}(z,s) \int d \omega \frac{C_\omega}{\tau_\omega} = \int d \omega \frac{C_\omega}{2 \Lambda_\omega\tau_\omega} \int_0^\infty dz' \left(\hat{h}(z',s)+ \frac{\tau_\omega \tilde{Q_\omega}(z')}{C_{\omega}T_0} \right)\left(F_1(|z-z'|) + 2F_2(z)F_2(z')\right)
\end{equation} 

This is an integral equation in the spatial variable $z$ for the non-dimensional temperature in the Laplace domain, which after solving, requires an inverse Laplace transform in order to obtain the full temperature solution in the time domain. For the thermal distribution, the spectral heat generation takes the form: 

\begin{equation}\label{eq:heatgenBTE}
\tilde{Q}_{\omega}(z) = \frac{C_\omega}{C}U_0\beta\mathrm{e}^{-\beta z}
\end{equation} 

Note that $\frac{C_\omega}{C}$ is a weighting of the contribution of a given mode to heat generation under the assumption of thermalized distribution and is different than the form found in~\cite{hua2014analytical}. While other distributions can be taken, we utilize this form in order to compare to the Fourier heat conduction solution.

\subsubsection{Variational solution}
Given the mathematical challenges in finding a closed solution to Eq.~\ref{eq:intBTE}, we opt for a simpler path. The insight is to take the known Fourier heat conduction solution (Eqs.~\ref{eq:fourierTemp},~\ref{eq:fourierTemp2}) as a starting point for the variational trial function. The simplest trial function is to take the diffusive temperature profile and allow just the thermal diffusivity to be a variational parameter. In general, the size effects exhibited by the BTE will affect both the temporal as well as the spatial distributions of the temperature. However, the simple variational solution that varies only one parameter, the thermal diffusivity, performs admirably by approximately solving for the thermal decay from the BTE over a broad range of grating period length scales. We proceed by taking the Fourier heat conduction solution of Eq.~\ref{eq:fourierTemp} as a trial function and use the thermal diffusivity as the variational parameter.

To solve for the variational parameter, we can utilize mathematical optimization methods such as least squares on the error residual of the temperature equation~\cite{chiloyan2016variational}, or impose a physical condition that we wish the trial function to satisfy. Here, we impose that the trial function must satisfy energy conservation taken over the control volume of the semi-infinite substrate over all time, analogous to the condition utilized for the thin film TTG geometry~\cite{chiloyan2016variational2}. This mathematical condition can be obtained by integrating the BTE of Eq.~\ref{eq:equilBTE} over the solid angle and frequency, and then also over the depth variable $z$ as well as over all time to yield:

\begin{equation}\label{eq:energyCons}
U_0 \frac{\lambda}{\pi} = 2i \int_0^\infty dz \int_0^\infty dt \tilde{q}_x(z,t)
\end{equation} 
 
This statement says that the total energy per unit area perpendicular to the $z$-axis deposited in the semi-infinite substrate initially (left hand side of Eq.~\ref{eq:heatgenBTE}) must be equal to the total energy that moves away in the in-plane direction. The in-plane heat flux is obtained by utilizing the spectral energy density of Eq.~\ref{eq:spectralBTE}, and integrating over the frequency and solid angle $ \hat{\tilde{q}}_x(z,s) = \int d\omega \int d\Omega \Theta v_\omega\mu_x\hat{\tilde{g}}_{\omega}(z,s,\mu_x,\mu_z)$ to obtain the in-plane heat flux: 
 
\begin{equation}\label{eq:fluxBTE}
\hat{\tilde{q}}_x(z,s)= \frac{T_0}{2}\int d \omega \frac{C_\omega v_\omega}{\Lambda_\omega} \int_0^\infty dz' \left(\hat{h}(z',s)+ \frac{\tau_\omega \tilde{Q_\omega}(z')}{C_{\omega}T_0} \right)\left(G_1(|z-z'|) + 2G_2(z)F_2(z')\right)
\end{equation} 
where we have defined the solid angle integral function:

\begin{equation}
G_n(z) = \frac{1}{2 \pi}\int d\Omega \Theta(\mu_z)\mu_z^{n-2}\mu_x\mathrm(e)^{-Vz}
\end{equation} 
 
Inserting the heat flux expression of Eq.~\ref{eq:fluxBTE} into the energy conservation statement of Eq.~\ref{eq:energyCons}, and inputting the variational trial function of the Fourier heat conduction solution of Eq.~\ref{eq:fourierTemp} as well as the thermal distribution for the heat generation rate, we can solve for the effective thermal conductivity after cleaning up some of the solid angle integrals. We obtain a form similar in structure to the results from the thin film TTG~\cite{chiloyan2016variational2} and the one-dimensional limit of the TTG~\cite{chiloyan2016variational}:

\begin{equation}
k = \frac{\frac{1}{3}\int d\omega C_\omega v_\omega \Lambda_\omega f(\eta_\omega,\mathrm{Kn}_\omega)}{\frac{1}{C}\int d\omega C_\omega g(\eta_\omega,\mathrm{Kn}_\omega)}
\end{equation} 
where $\mathrm{Kn}_\omega = \Lambda_\omega \beta$ and $f$ and $g$ are the kernels that weigh a given mode's contribution to effective thermal conductivity under the imposed size effects, explicitly given as

\begin{equation}
\begin{split}
f(\eta_\omega,\mathrm{Kn}_\omega) &=\frac{3}{\eta_\omega^2}\left(1-\frac{1}{\eta_\omega}\mathrm{arctan}(\eta_\omega) + \frac{\eta_\omega^2 \Psi(\eta_\omega,\mathrm{Kn}_\omega)- \mathrm{Kn}_\omega^2 \Psi(\eta_\omega,\mathrm{Kn}_\omega)}{\eta_\omega^2-\mathrm{Kn}_\omega^2}\right)\\
g(\eta_\omega,\mathrm{Kn}_\omega) &=\frac{1}{\eta_\omega}\mathrm{arctan}(\eta_\omega) + \Psi(\eta_\omega,\mathrm{Kn}_\omega)\\
\end{split}
\end{equation}
We have defined the following solid angle integral functions: 

\begin{equation}
\begin{split}
\Psi(x,z) &= \frac{1}{2}\psi_1(x,z)- \frac{1}{1+\sqrt{1+x^2}}\psi_0(x,z) \\
\psi_n(x,z) &= \frac{1}{2\pi} \int d \Omega \Theta(\mu_z)\frac{z \mu_z}{(1+ix\mu_x)^n (1+z\mu_z +ix\mu_x)}
\end{split}
\end{equation} 
 
If we take the limit of $\mathrm{Kn}_\omega \to 0$, i.e. the case of very long penetration depth, the solid angle integrals vanish as $\psi_n (\eta_\omega,\mathrm{Kn}_\omega \to 0) \propto \mathrm{Kn}_\omega$, and we recover the one-dimensional TTG limit as in this case the substrate essentially starts off at a uniform temperature, and we recover the previously derived effective thermal conductivity~\cite{chiloyan2016variational}. Note that information concerning the spectral contribution to heat capacity in needed, unlike prior work~\cite{yang2013mean}, in the equation for effective thermal conductivity. The more interesting case for this problem is the reduction to surface heating, i.e. $\mathrm{Kn}_\omega \to \infty$. In this case, the kernel functions simplify to:

\begin{equation}
\begin{split}
f(\eta_\omega,\mathrm{Kn}_\omega \to \infty) &=\frac{3}{2\eta_\omega^2}\left(1-\frac{1}{\eta_\omega}\mathrm{arctan}(\eta_\omega)\right) -\frac{1}{\eta^3(1+\sqrt{1+\eta^2})}\left((1+\eta_\omega)^{\frac{3}{2}}-\frac{3}{2}\eta_\omega^2-\eta_\omega^3-1\right)\\
g(\eta_\omega,\mathrm{Kn}_\omega \to \infty) &=\frac{1}{2\eta_\omega}\mathrm{arctan}(\eta_\omega)+\frac{1}{1+\sqrt{1+\eta^2}}
\end{split}
\end{equation} 
 
For the general case of arbitrary penetration depth, the solid angle integral functions can be calculated analytically, which allows for a fully analytical effective thermal conductivity for any penetration depth into the substrate.

\subsubsection{Comparison between the Variational Solution and Monte Carlo Simulations}
\begin{figure}
\centering     %%% not \center
\subfloat[]{\label{fig:a}\includegraphics[width=0.45\textwidth]{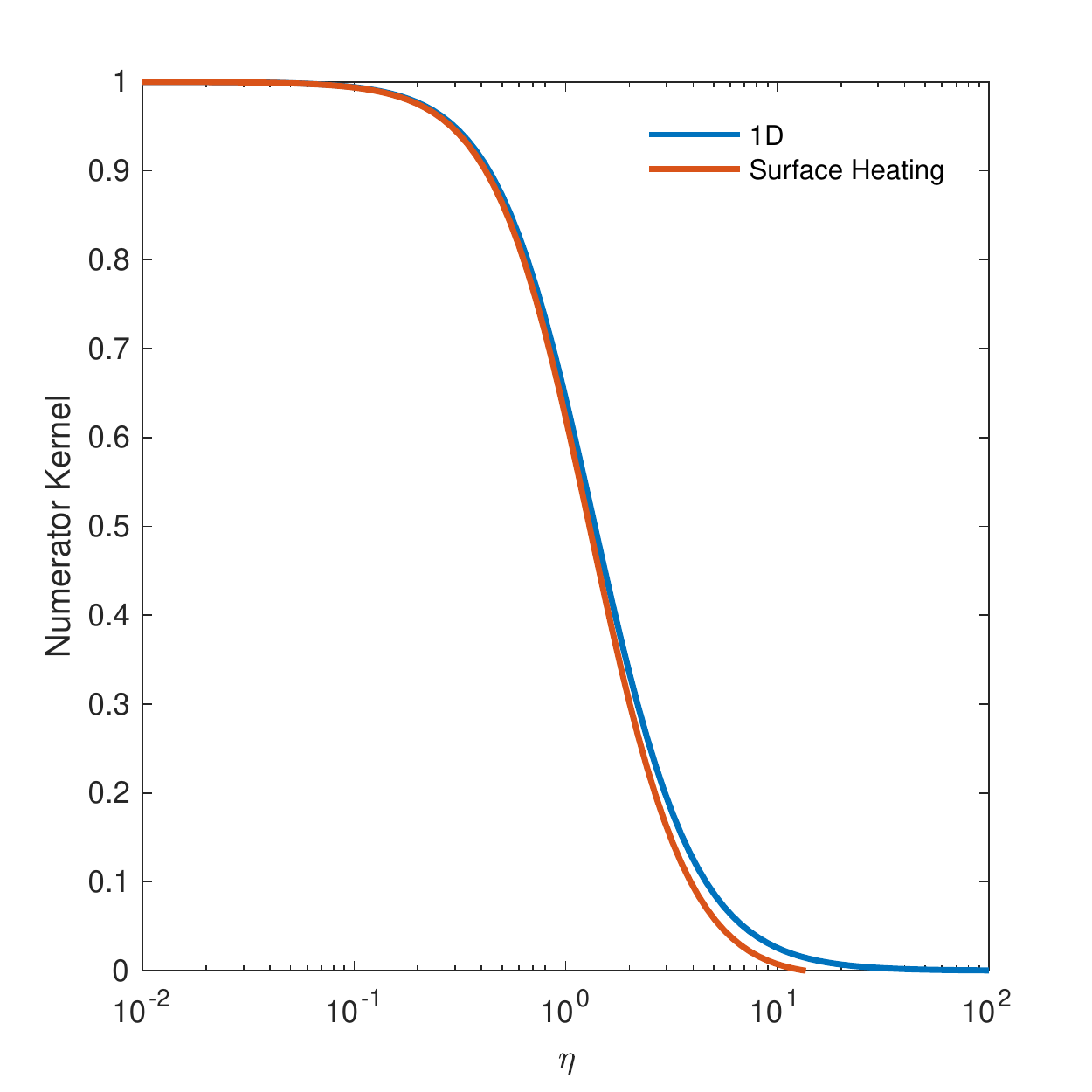}}
\subfloat[]{\label{fig:b}\includegraphics[width=0.45\textwidth]{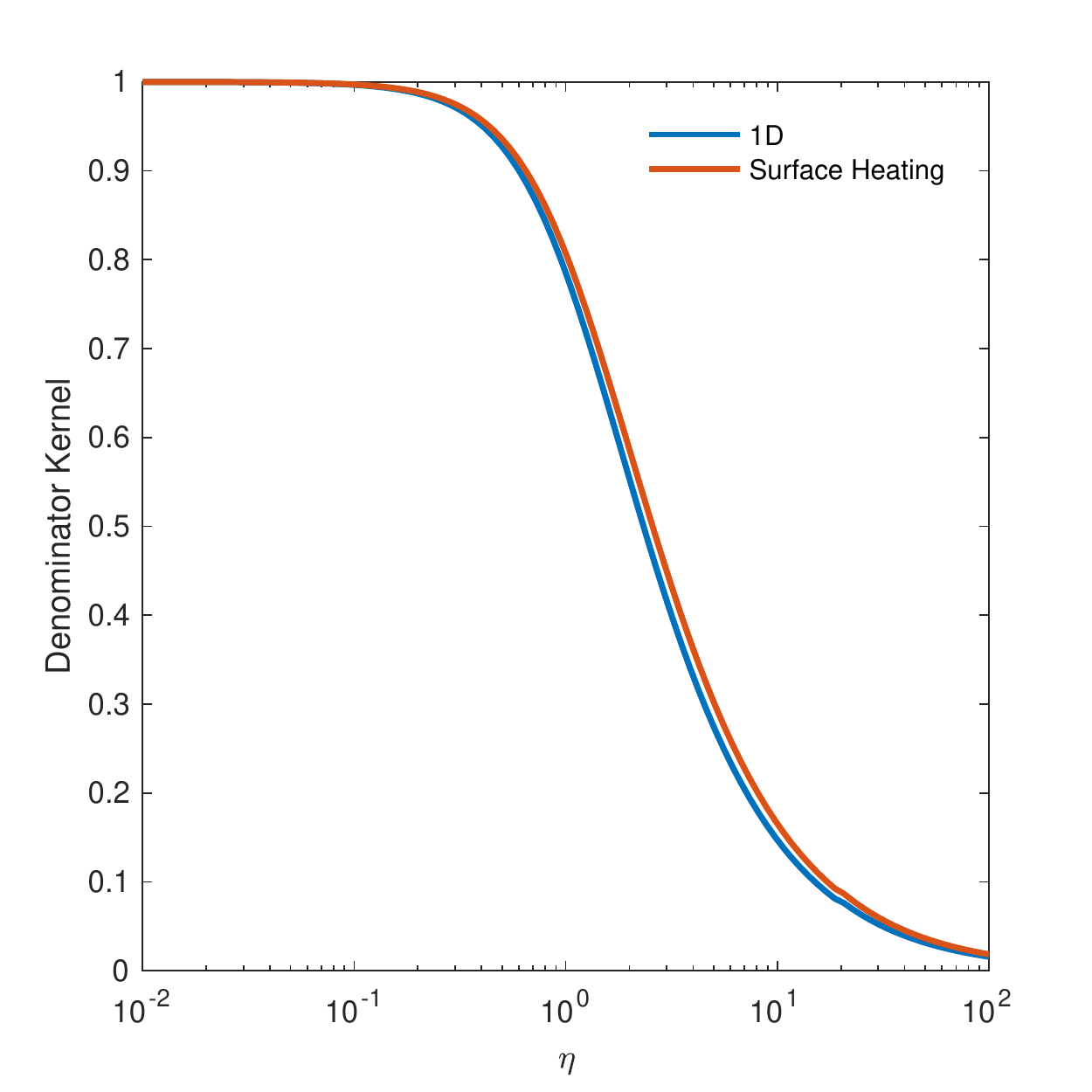}}
\caption{Kernels of the effective conductivity. The numerator kernel $f$ (a) that shows the size effects and appears beside the differential conductivity and the denominator kernel $g$ (b) that shows the size effects and appears beside the spectral heat capacity. The optical penetration depth does not have a large effect on the effective thermal conductivity with the diffuse surface boundary condition.}\label{fig:kernels}
\end{figure}

To study the effect of the optical penetration depth in the case of a diffuse surface boundary condition, we first plot the kernels $f$ and $g$ as a function of $\eta$ for the extremal limits of $\mathrm{Kn}_\omega$. The one dimensional limit of $\mathrm{Kn}_\omega \to 0$ and the surface heating limit of $ \mathrm{Kn}_\omega \to \infty$ define the envelope of curves for which the kernels for arbitrary values of the penetration depth must lie between. As the Knudsen number increases, the size effect due to the optical penetration depth increases, which physically results in a decrease of the effective thermal conductivity. This occurs due to the decrease in the numerator kernel $f$, and the increase of the denominator kernel $g$. Figure~\ref{fig:kernels} shows that one dimensional limit and the surface heating limit are practically indistinguishable, indicating that the effective thermal conductivity due to a diffuse boundary experiences weak effects from the optical penetration depth.

\begin{figure}
\centering     %%% not \center
\includegraphics[width=0.66\textwidth]{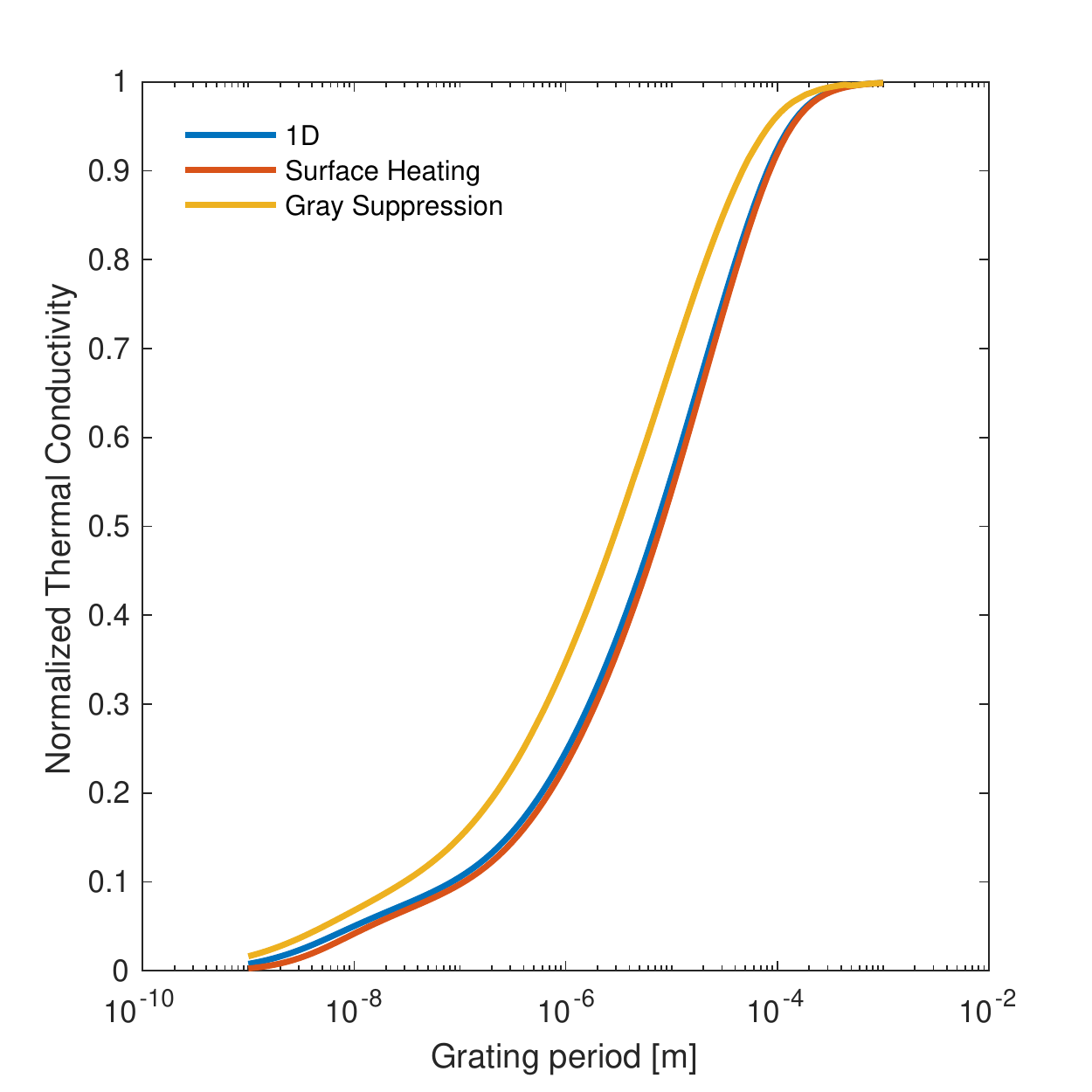}
\caption{Effective thermal conductivity for Si$_{93.4}$Ge$_{6.6}$ in the one-dimensional limit and the surface heating limit. The effective thermal conductivity using the gray suppression function for one-dimensional TTG from~\cite{chiloyan2016variational} is also shown.}\label{fig:effK}
\end{figure}

Utilizing the derived kernels to calculate the effective thermal conductivity for Si$_{93.4}$Ge$_{6.6}$, we show in Figure~\ref{fig:effK} the effective thermal conductivity in the various limits. Note that the effective thermal conductivity is quite similar in the one dimensional limit and in the surface heating limit. As expected, when the thermal grating period is much smaller than the optical penetration depth, the effective thermal conductivity takes on values of the one dimensional limit, as the transport is mostly in-plane. In the opposite case, when the grating period is much larger than the optical penetration depth, the effective thermal conductivity limits to the surface heating limit.

Figure~\ref{fig:effK} demonstrates that the variational technique predicts that transport has a weak dependence on the optical penetration depth, a consequence of the kernels' weak dependence on optical penetration depth. As such, the one-dimensional limit of the TTG~\cite{chiloyan2016variational} geometry is sufficient to characterize the dependence of effective thermal conductivity on grating period.

\begin{figure}
\centering     %%% not \center
\subfloat[]{\label{fig:a}\includegraphics[width=0.45\textwidth]{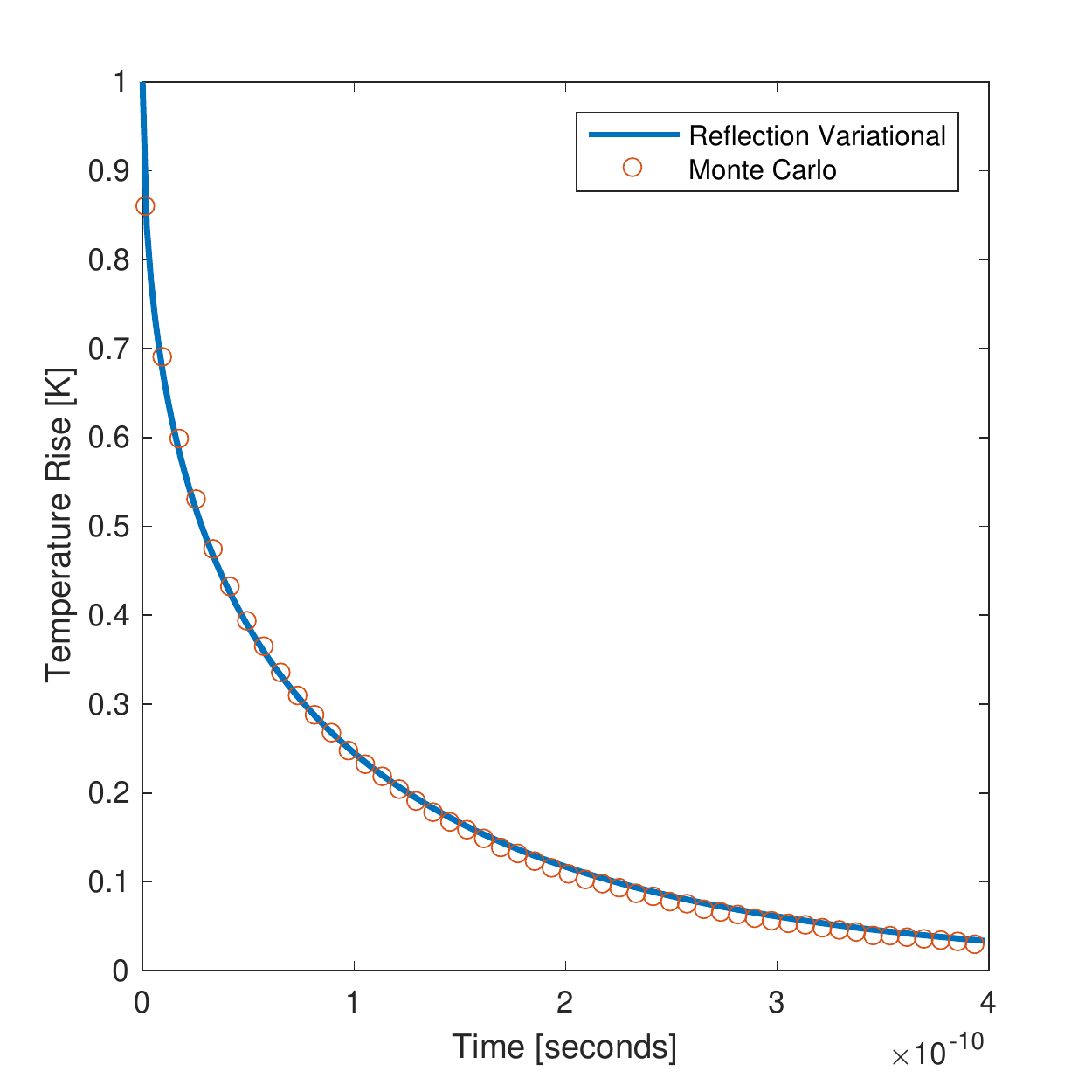}}
\subfloat[]{\label{fig:b}\includegraphics[width=0.45\textwidth]{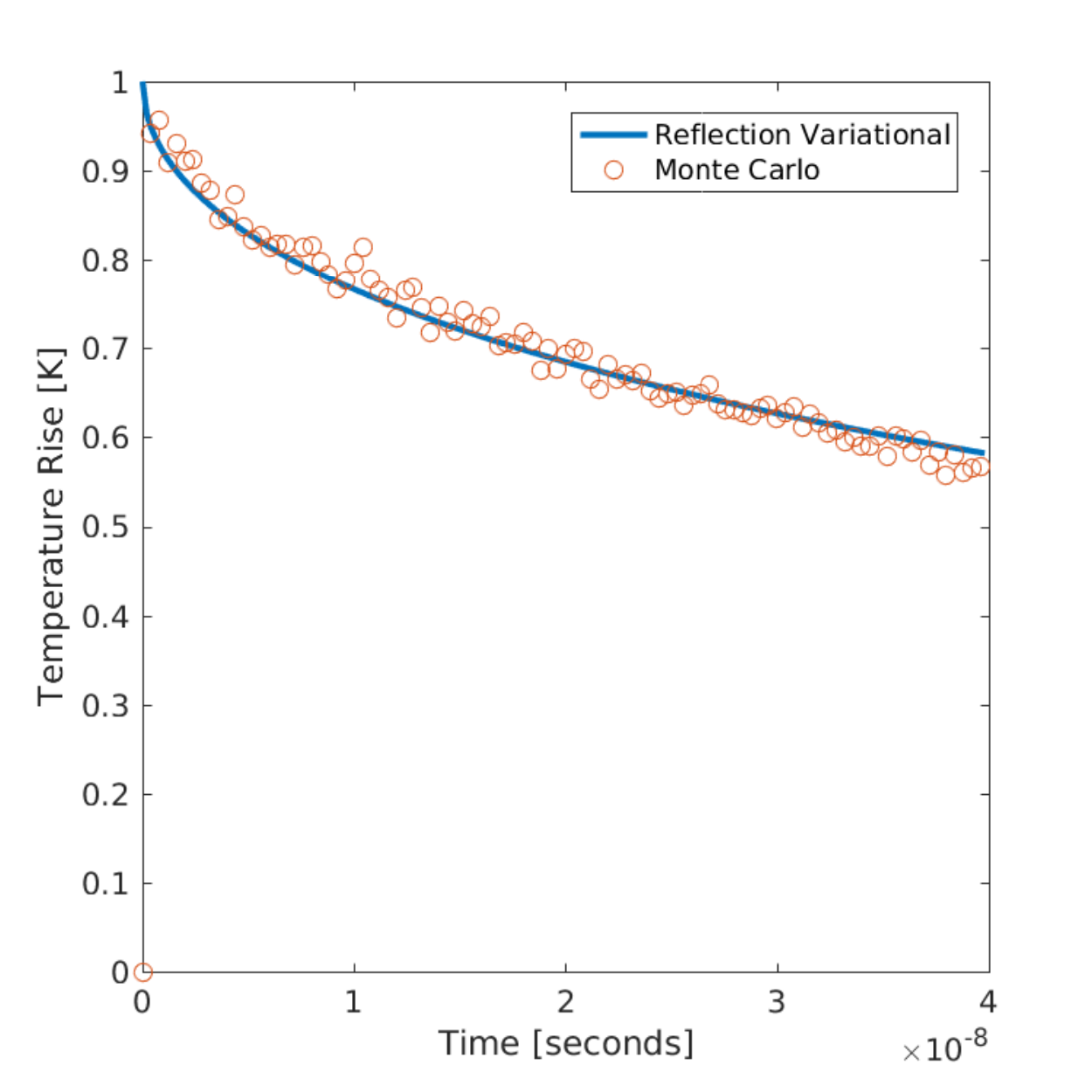}}
\caption{Temperature profiles obtained from Monte Carlo simulations compared with the corresponding variational predictions for Si$_{93.4}$Ge$_{6.6}$ at 300 K with a (a) grating period of 100 nm and optical penetration depth of 10 nm and a (b) grating period of 10 um and optical penetration depth of 1 um. The Monte Carlo trace for case (b) contains noise because of the computational cost of simulating longer decays for a large number of effective particles.}\label{fig:mc_v_var}
\end{figure}

In the limit of large grating periods, the thin film TTG limits to the Fuchs-Sondheimer~\cite{sondheimer1952mean} problem of in-plane transport, and there is still a reduction of the effective thermal conductivity due to the finite size of the membrane. In contrast, for the reflection TTG, the limit of large grating period yields the bulk thermal conductivity, regardless of the optical penetration depth. Thus a modified Fourier approach will fail to capture the details of a thermal decay due to a localized heat source (i.e. delta function in space and time in a semi-infinite geometry). In this case, the transport at short times (on the order of the dominant relaxation times) is initially ballistic and given sufficient scattering, the transport becomes diffusive. The variational method, using the Fourier temperature profile as input, reveals that the thermal conductivity that best recovers this behavior is bulk. This can be understood as a consequence of the constraint imposed by the equilibrium condition of Eq.~\ref{eq:energyCons}, which dictates the behavior of the variational temperature profile in the large time limit where transport is diffusive. An example of this limitation is presented in the supplementary material. To ensure that this limitation is not present in the current experimental study, we compare against established Monte Carlo simulations of the RTA-BTE~\cite{peraud2011efficient,peraud2012alternative}. As is seen in Figure~\ref{fig:mc_v_var}, agreement at a grating period of 100 nm and an optical penetration depth of 10 nm and for a grating period of 10 um and an optical penetration depth of 1 um is observed.

% \begin{figure}
% \centering     %%% not \center
% \includegraphics[width=0.66\textwidth]{{MC_v_Var.SiGe.10nmPD.100nmL}.pdf}
% \caption{Temperature profile obtained from a Monte Carlo simulation for Si$_{93.4}$Ge$_{6.6}$ with a grating period of 100 nm and optical penetration depth of 10 nm compared with the corresponding variational prediction.}\label{fig:mc_v_var}
% \end{figure}

As our experimental conditions operate at penetration depths on the order of 1 um for Si$_{93.4}$Ge$_{6.6}$ \cite{braunstein1958intrinsic}, with grating periods of between 1 and 13.5 um, by the pigeonhole principle, we can move forward with our variational solutions.

\section{Experiment \label{sec:experiment}}
\subsection{Sample specifications}
The deposition of the SiGe sample was done by metal-organic chemical vapor deposition (MOCVD). Briefly, SiH$_4$ and GeH$_4$ enter the reactor, which break up into Si, Ge, and H$_2$ from exposure to high temperatures (750-800C). The composition is controlled by tunning the flow rates of SiH$_4$ and GeH$_4$. A single crystal sample consisting of 93.4\% Si, 6.6\% Ge with a thickness of 6 um on a Si wafer was used for this work.

\subsection{Transient Thermal Grating}

Transient grating spectroscopy is a variant on four-wave-mixing spectroscopic techniques that can measure thermal transport dynamics over a well-defined in-plane length scale. In this technique, two pump laser pulses (515 nm, 60 ps FWHM) are crossed at the surface of the sample, where they interfere to yield a sinusoidal intensity pattern. Absorption by the sample creates a matching temperature profile, which evolves as a function of time through in-plane and cross-plane transport. The time dynamics of this ``transient grating'' are measured by the diffraction of a quasi-continuous probe beam (532 nm), and phase-specific information is extracted through heterodyned detection of the TTG signal by superposition of the diffracted signal with a reference beam (local oscillator) derived from the probe beam source. The signal is detected using a fast photodiode (Hamamatsu C5658, 1 GHz bandwidth) and recorded on an oscilloscope (Tektronix TDS 7404, 4GHz bandwidth). Specific details of the optical setup can be found elsewhere \cite{maznev1998optical,johnson2012phase,vega2015laser}.

The TTG signal will in principle have both real and imaginary field contributions due to ``amplitude-grating'' and ``phase-grating'' responses, respectively. The phase grating contributions contains decay components that correspond to thermal expansion and the imaginary part of the thermoreflectance and acoustic oscillations corresponding to the impulsive stimulation of surface acoustic waves (SAWs), whereas the amplitude-grating response only contains one term corresponding to the real part of the thermoreflectance \cite{maznev1998optical}. Analysis of the amplitude-grating contribution is simpler due to the single contribution, and so this term was isolated during the measurements by optimizing the heterodyne phase to minimize the SAW signal which only appears in the phase-grating response.

\subsection{Results}

\begin{figure}[H]
\centering     %%% not \center
\includegraphics[width=0.66\textwidth]{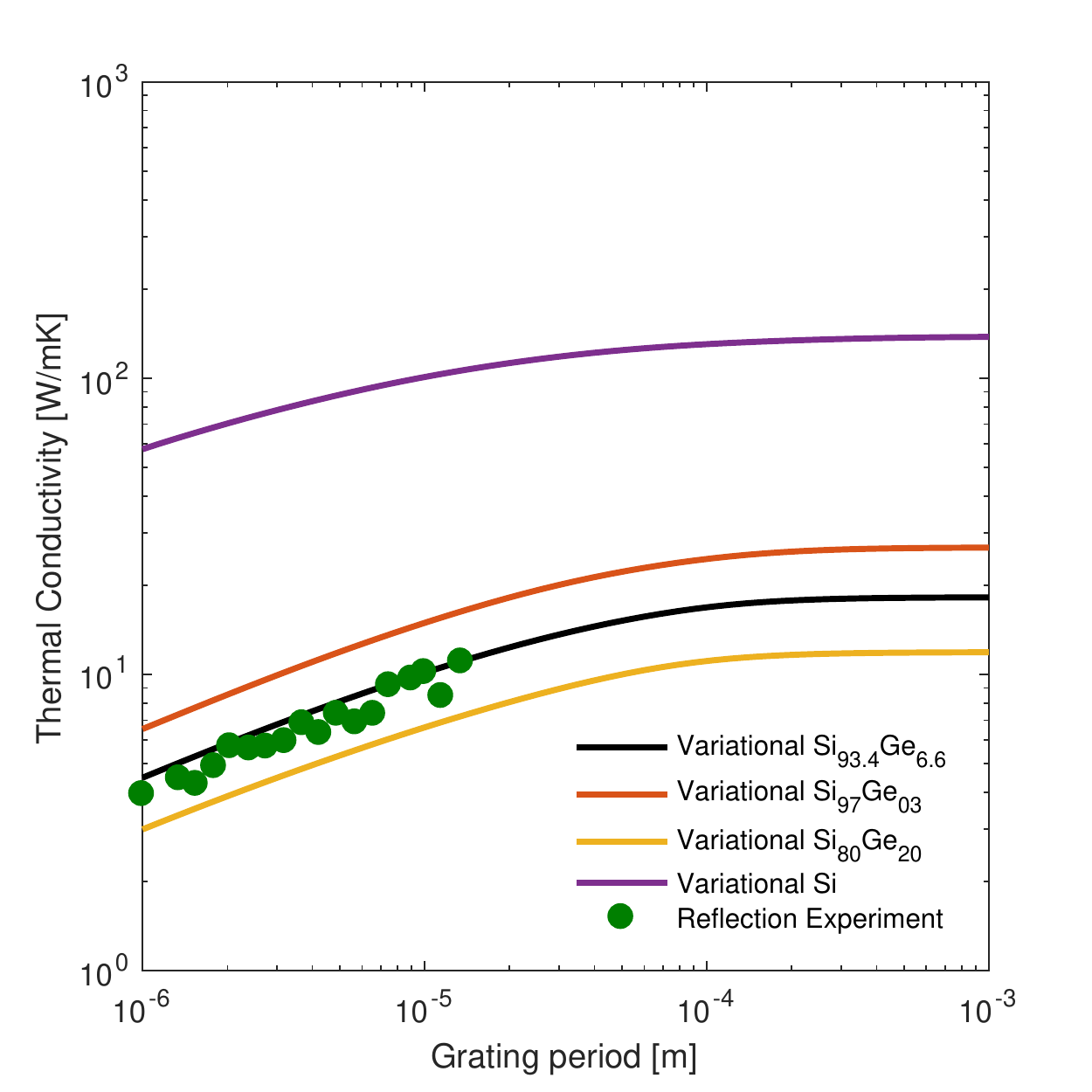}
\caption{Green circles correspond to measured TTG data for a range of grating periods, from 13.5 to 1 um. The black line is the prediction from the variational solution with DFT properties as input, while the orange line (yellow line, purple line) corresponds to the variational prediction for Si$_{97}$Ge$_{3}$ (Si$_{80}$Ge$_{20}$, Si).}\label{fig:ttg_results}
\end{figure}

All measurements of the Si$_{93.4}$Ge$_{6.6}$ sample were conducted at room temperature. Figure~\ref{fig:ttg_results} depicts the TTG measurements alongside the prediction from the variational solution using properties obtained from first principle calculations following Section~\ref{sec:theory}. We have used an optical penetration depth of 1500 nm, according to \cite{braunstein1958intrinsic}. The effect of uncertainty in the penetration depth is discussed in the supplementary material. The agreement is remarkable, considering we are simply fitting the TTG measurements to the Fourier-based temperature profiles (Eq.~\ref{eq:fourierTempSurf}) to extract effective thermal conductivity. This agreement persists for a range of grating periods, from $\sim$ 13.5 to 1 um. Example fits of the TTG data with comparisons to the variational predictions are found in the supplementary material.

\section{Discussion and Outlook}\label{sec:discussion}

Before delving into the nuances of the work, a quick reminder of what we have done. We calculated the first principles phonon properties to match the exact composition of the sample studied experimentally. We then used these properties and the variational solution to the BTE to predict (without any fitting required) the recorded observable of TTG experiments, the temperature decay. In doing so, we uncover excellent agreement between the effective thermal conductivities of theory and experiment.

\begin{figure}
\centering     %%% not \center
\includegraphics[width=0.66\textwidth]{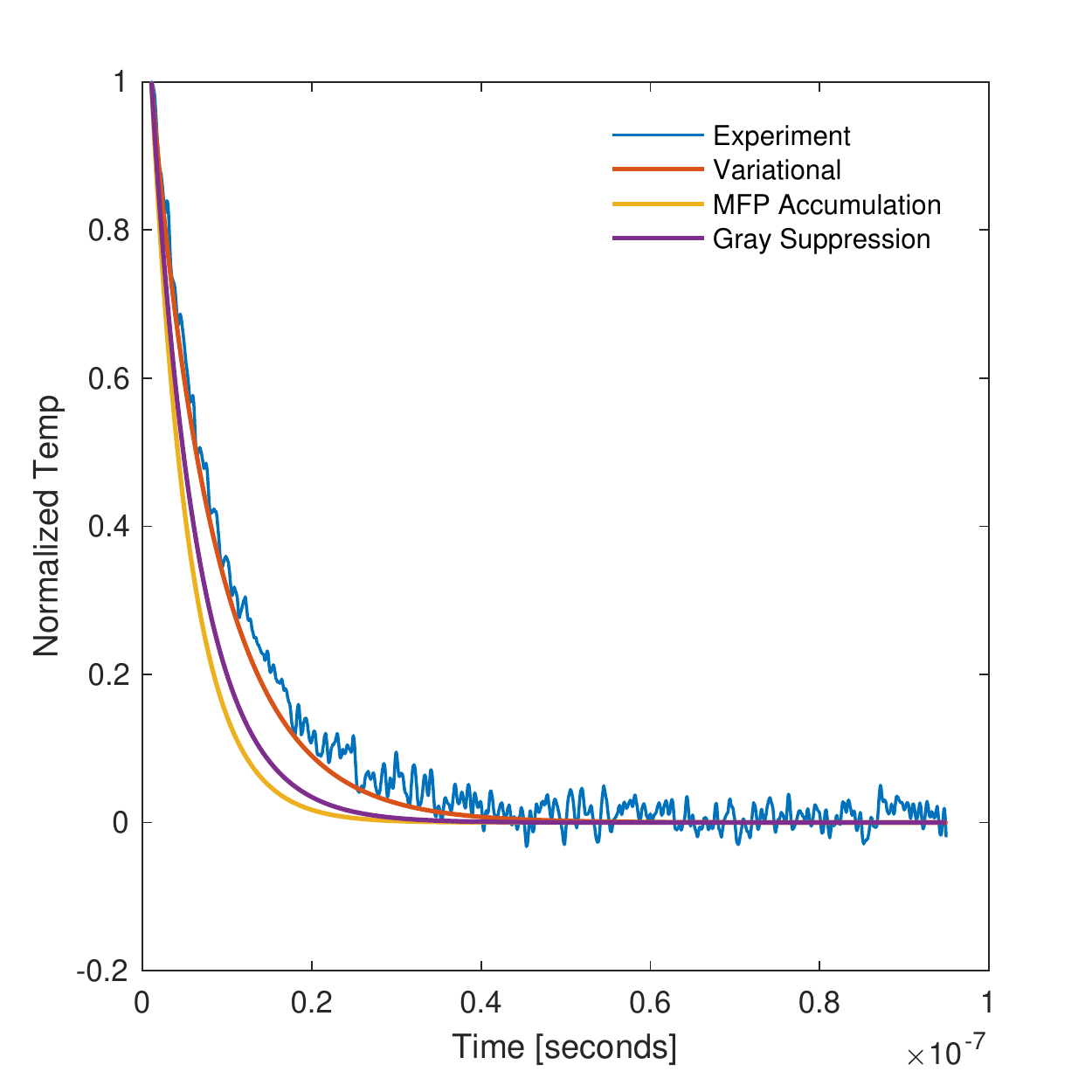}
\caption{Comparison between the predictions from the variational approach to using the MFP accumulation function or the gray suppression function for one-dimensional TTG from~\cite{chiloyan2016variational} to estimate the effective thermal conductivity at 1.00 um grating period.}\label{fig:ttg_MFPvVar}
\end{figure}

One of the first explanations of size effects in SiGe grew out of the observation of frequency dependence in TDTR measurements \cite{koh2007frequency}. This explanation relied on the application of thermal penetration depth, $L_{tpd}\sim\sqrt{\frac{\alpha_{bulk}}{\omega}}$, as a heuristic approximation to estimate the magnitude of the deviation from a bulk thermal conductivity. For Si$_{93.4}$Ge$_{6.6}$, $\alpha_{bulk} = 1.2358\text{E-5} ~\text{ m}^2\text{/s}$, with 10 MHz, yields a $L_{tpd} \sim 1$ um. Under this approximation, we can take the MFP thermal conductivity accumulation function at 1 um, yielding $0.4 k_{bulk} = 7.3$ W/mK. From our TTG results, using $L_{tpd}~\sim\lambda$, we find $0.25 k_{bulk} = 4.5$ W/mK. By this same argument, frequency dependence should also be observed in silicon with $\alpha_{bulk} = 8.8\text{E-5}~\text{ m}^2\text{/s}$, which at 10 MHz, yields a $L_{tpd} \sim 3$ um and $\sim 0.7k_{bulk} = 98$ W/mK from the MFP accumulation function~\cite{esfarjani2011heat}, but $k_{exp} > 120 $ W/mK for the same frequency range is often reported~\cite{wilson2014anisotropic,wilson2015limits}. The reason for this discrepancy has not been satisfactorily resolved \cite{ding2014radial,vermeersch2016limitations}. For example, the results of Hua et al.~\cite{hua2016experimental} and Wilson et al.~\cite{wilson2014anisotropic} suggest that the reported thermal conductivity obtained from a TDTR measurement is dependent upon the interface conductance, indicating that this thermal conductivity can no longer be interpreted as an intrinsic property of the material. Meanwhile, the penetration depth argument has been used to interpret frequency dependence in BB-FDTR measurements \cite{regner2013broadband}, suggesting that this tool could be used for phonon MFP spectroscopy. The next natural step in the interpretations of deviations from bulk required theory to move beyond the Heaviside cutoff of the thermal penetration depth and obtain a gray suppression function from solving the gray BTE \cite{minnich2012determining,regner2014analytical,zeng2014measuring}. This function is then used as a kernel in the effective thermal conductivity integral. This picture has also turned out to be an oversimplification \cite{collins2013non, chiloyan2016variational}. Here we show that a fully spectral solution to the BTE is required to characterize the effective conductivity. This progression from penetration depth to gray suppression to fully spectral interpretations is shown in Figure~\ref{fig:ttg_MFPvVar}. 

%Given the experimental challenges of observing size effects on thermal transport, we stress the distinction between explanation and prediction. An explanation leads to \textit{a posterior} expectation and a prediction leads to \textit{a priori} expectation. While the difference may be subtle, for experimental techniques to be useful, we require the latter. Earlier attempts were only capable of explaining the frequency dependence of TDTR~\cite{koh2007frequency,da2012role,vermeersch2015superdiffusive2}. In the case of TDTR, without a spectral BTE-based description, we can only attempt to explain the results obtained (and it is challenging to do so in a consistent manner), but have so not been capable of predicting those results. This is insufficient if we are to use TDTR as a platform for MFP reconstruction~\cite{cuffe2015reconstructing}.

%While not free of its own disadvantages, TTG does not appear to be burdened in terms of complexity in the way that TDTR is. 

In contrast to the interpretation of thermal penetration depth of TDTR, the length scales in TTG do not depend on the intrinsic value of a material's transport coefficient, and are therefore physically well-defined independent variables. Although the information concerning the optical penetration depth is required, this is well within current characterization technology~\cite{fuyuki2005photographic}. Given that the variational solutions to the 1D and surface heating TTG geometries predict approximately the same effective thermal conductivity dependence on grating period, we have theoretical bounds on the observed experimental decay curves from which the the transport regime can be determined (i.e. purely 1D, finite penetration depth, or surface heating). In doing so, we have presented a theoretical framework that is falsifiable, given that experimental deviations from theory can be understood as departures from the approximations used in this work: the VCA, the RTA-BTE and the specific trial solution for the temperature profile used in the variational method. These approximations can be lifted, as will be shown in future works. With the methodology presented here, the TTG can be used to study in-plane transport in opaque thin films that require a supporting substrate.

As TDTR measurements are sensitive to the cross-plane transport, the TTG provides a complementary tool for measuring in-plane transport. The variational method can be extended to more complicated geometries, such as layered systems with interfaces, ideally suited for providing insight into the interpretations of TDTR and TTG measurements. Such an extension would provide a path towards unifiying the interpretations of the measurements from TDTR and TTG.

%As an example, the variational solution can be extended to use a trial solution with anisotropic diffusivity to account for strong cross-plane effects in the reflection mode geometry. Furthermore, we can generalize this methodology to materials where the RTA fails to describe the phonon transport, such a graphite and graphene, which will be subject to a future study.

%Vermeersch et al. noted that a one-dimensional TTG experiment contains a single spatial frequency and the decay is near-exponential, suggesting that a Fourier based interpretation is valid, and on the other hand, TDTR is a pulse in time and space and therefore contains many spatial frequencies~\cite{vermeersch2015superdiffusive,vermeersch2015superdiffusive2}. They proceed to demonstrate that superdiffusive behavior can explain the thermal transport in alloys in TDTR-like experimental geometries. In the reflection TTG geometry, the temperature decay is not exponential and contains a cross-plane effect due to the finite optical penetration depth, however, no invocation of superdiffusive statistics is necessary to capture the temperature decays for Si$_{93.4}$Ge$_{6.6}$ in this geometry for these length scales. 

\section{Conclusion \label{sec:conclusion}}

Our TTG experimental results augmented with DFT-based modeling and the variational BTE solution indicate that this experimental geometry is capable of meeting the predictive criteria necessary for studying size effects on thermal transport in complex materials, such as the SiGe alloy studied here. Interesting questions can now be asked, such as in what systems or at what length scales can we expect to find a breakdown of the VCA. This geometry will likely prove useful in the study of systems where the relaxation time approximation fails. The TTG experiment provides a path towards tabletop studies of the microscopic properties of thermal transport.

\section{Acknowledgments}

We acknowledge Jiawei Zhou and Bai Song for helpful discussions. This work was done as part of the Solid-State Solar-Thermal Energy Conversion Center (S3TEC)  an Energy Frontier Research Center funded by the U.S. Department of Energy (DOE), Office of Science, Basic Energy Sciences (BES), under Award DE-SC0001299 / DE-FG02-09ER46577.

\bibliography{references}
\newpage

\section{Appendix A}

\begin{figure}[H]
\captionsetup[subfloat]{farskip=1pt,captionskip=1pt}
\centering     %%% not \center
\subfloat[13.5 um]{\label{fig:a}\includegraphics[width=0.25\textwidth]{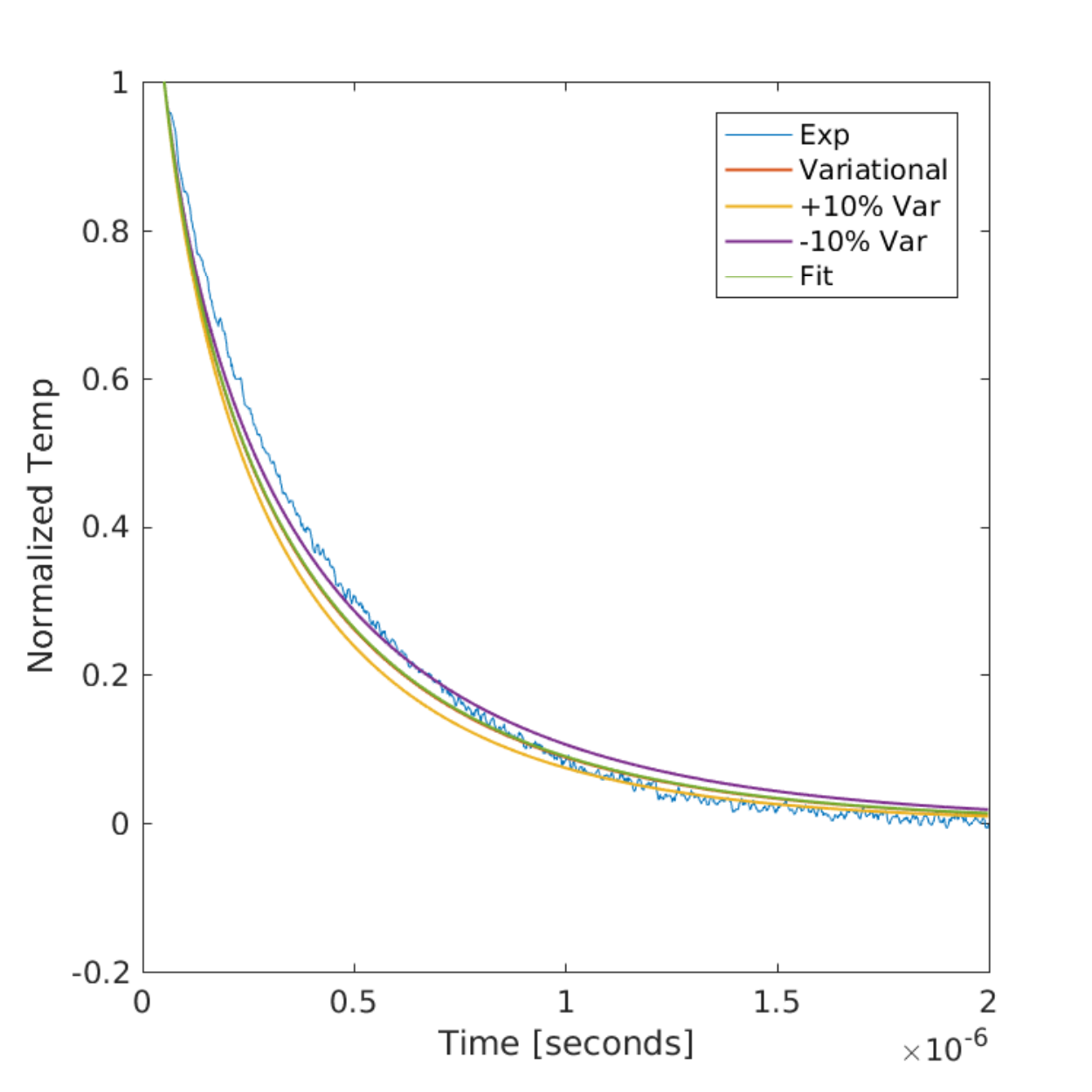}}
\subfloat[10.0 um]{\label{fig:b}\includegraphics[width=0.25\textwidth]{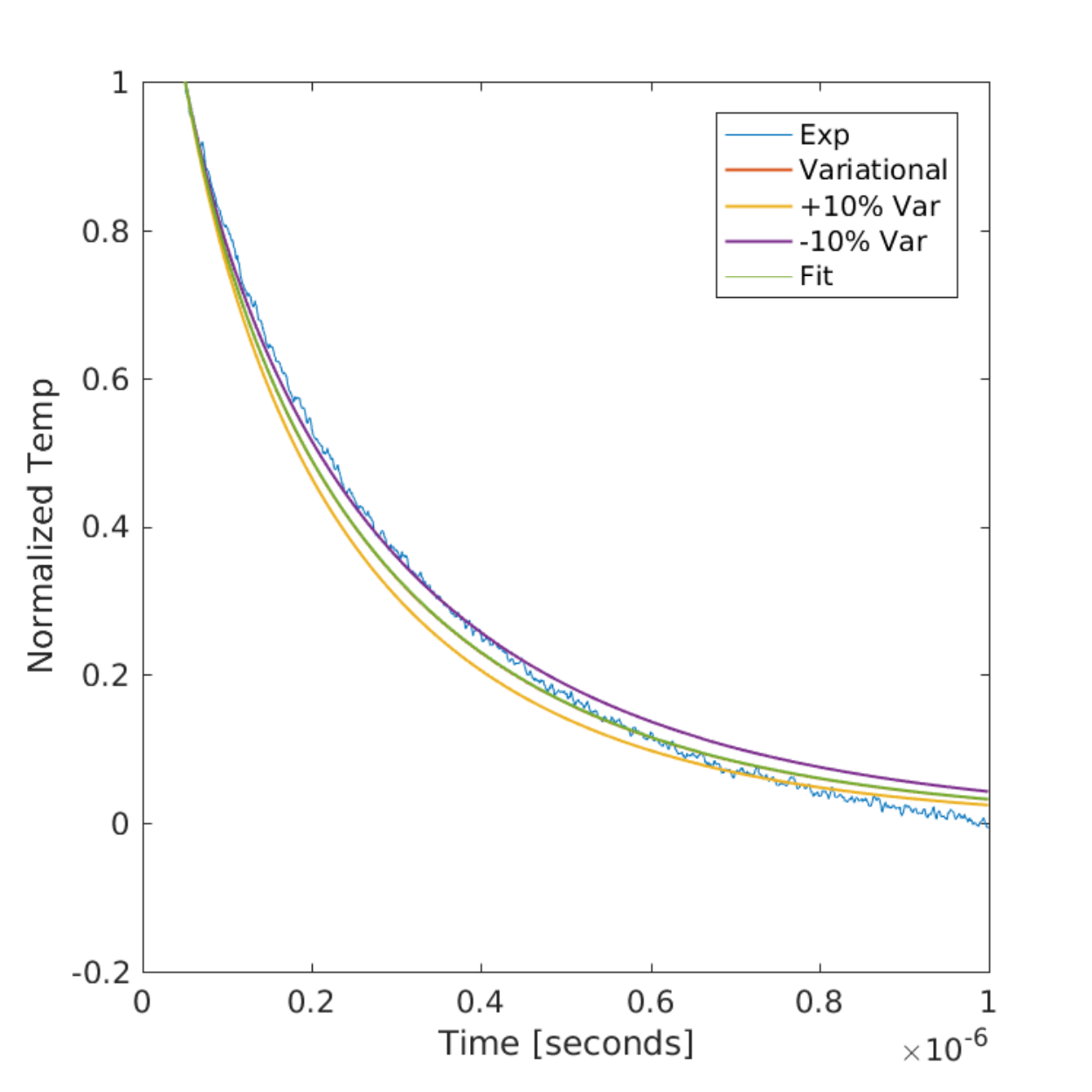}}
\subfloat[9.0 um]{\label{fig:a}\includegraphics[width=0.25\textwidth]{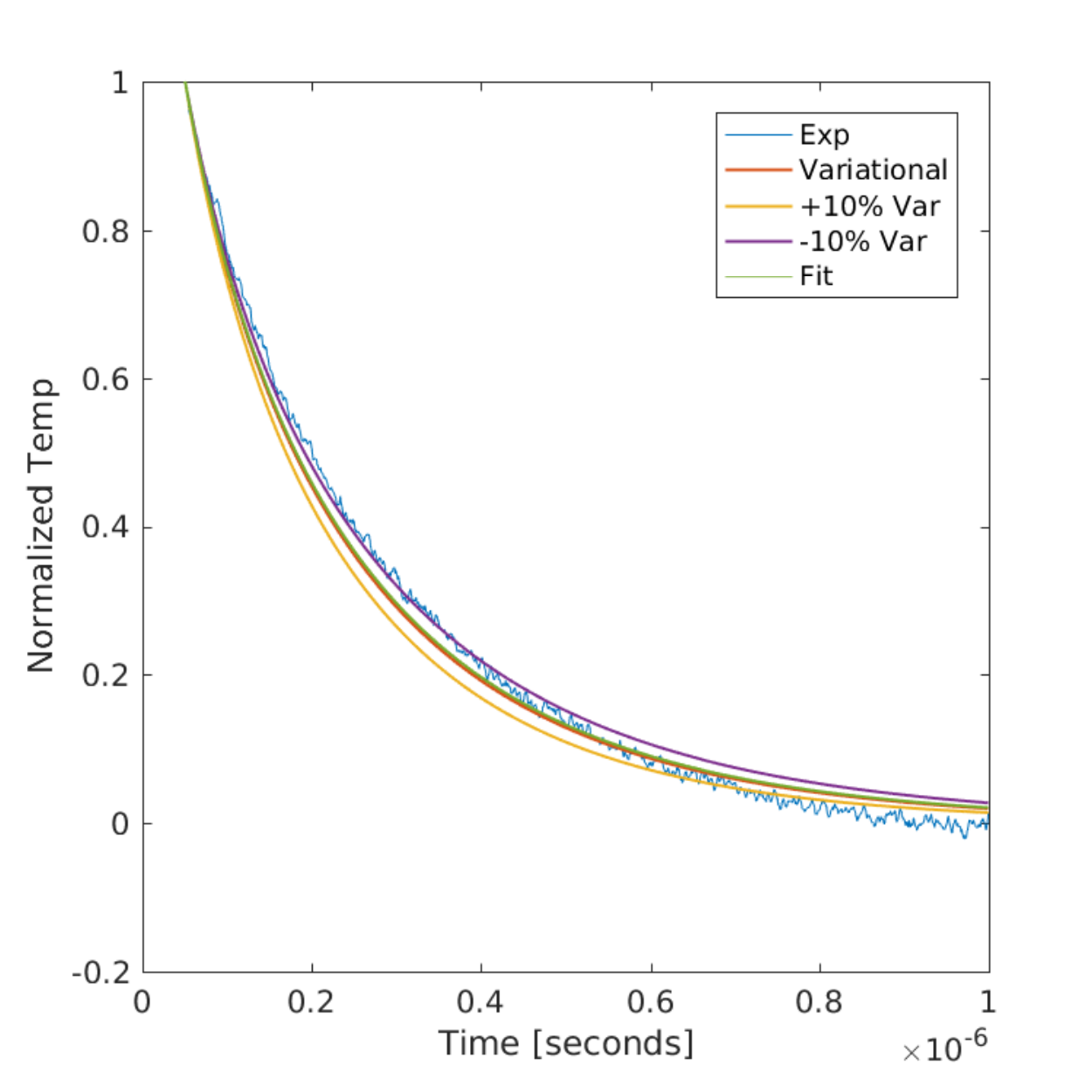}}
\subfloat[7.5 um]{\label{fig:b}\includegraphics[width=0.25\textwidth]{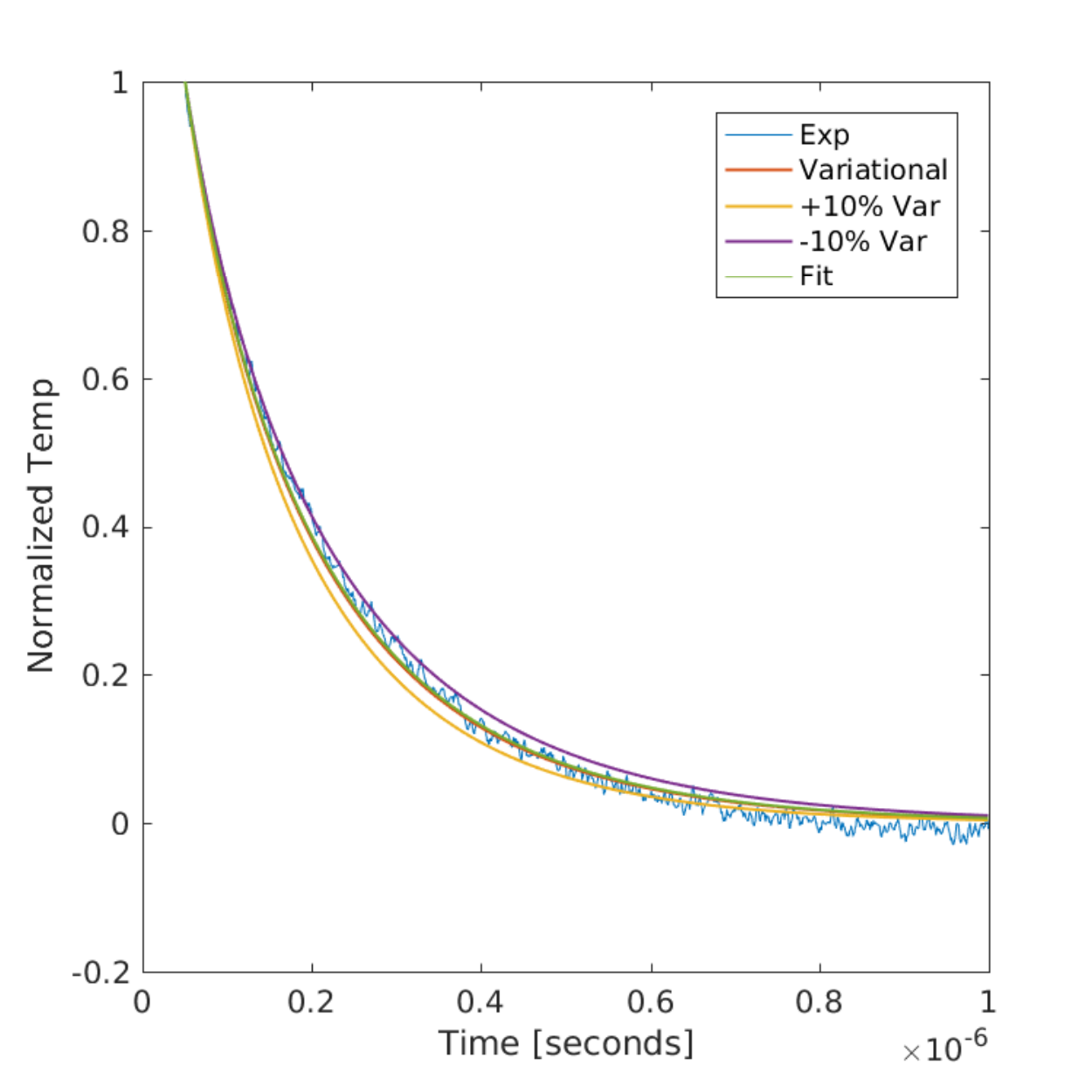}}
\hfill
\subfloat[6.6 um]{\label{fig:a}\includegraphics[width=0.25\textwidth]{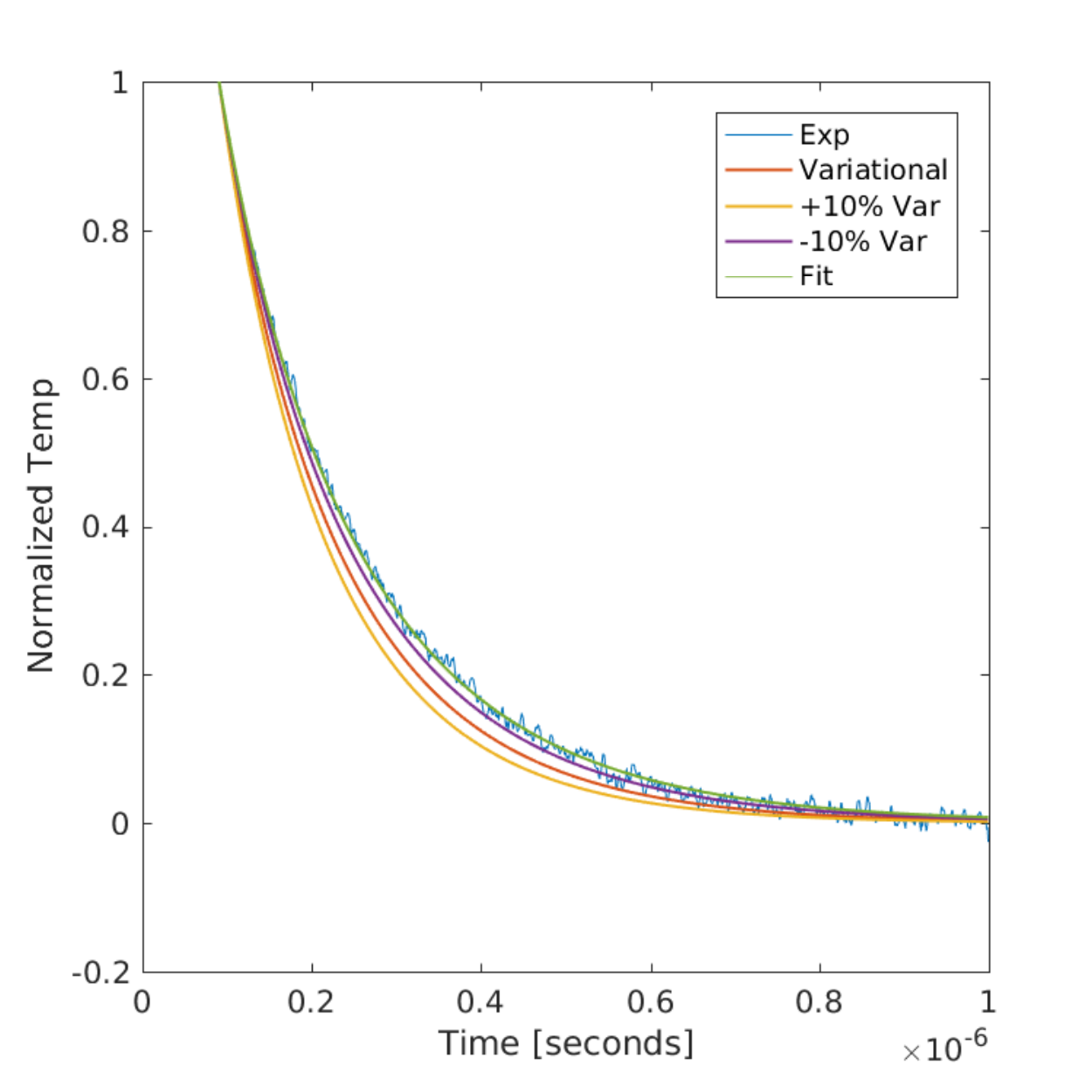}}
\subfloat[5.7 um]{\label{fig:a}\includegraphics[width=0.25\textwidth]{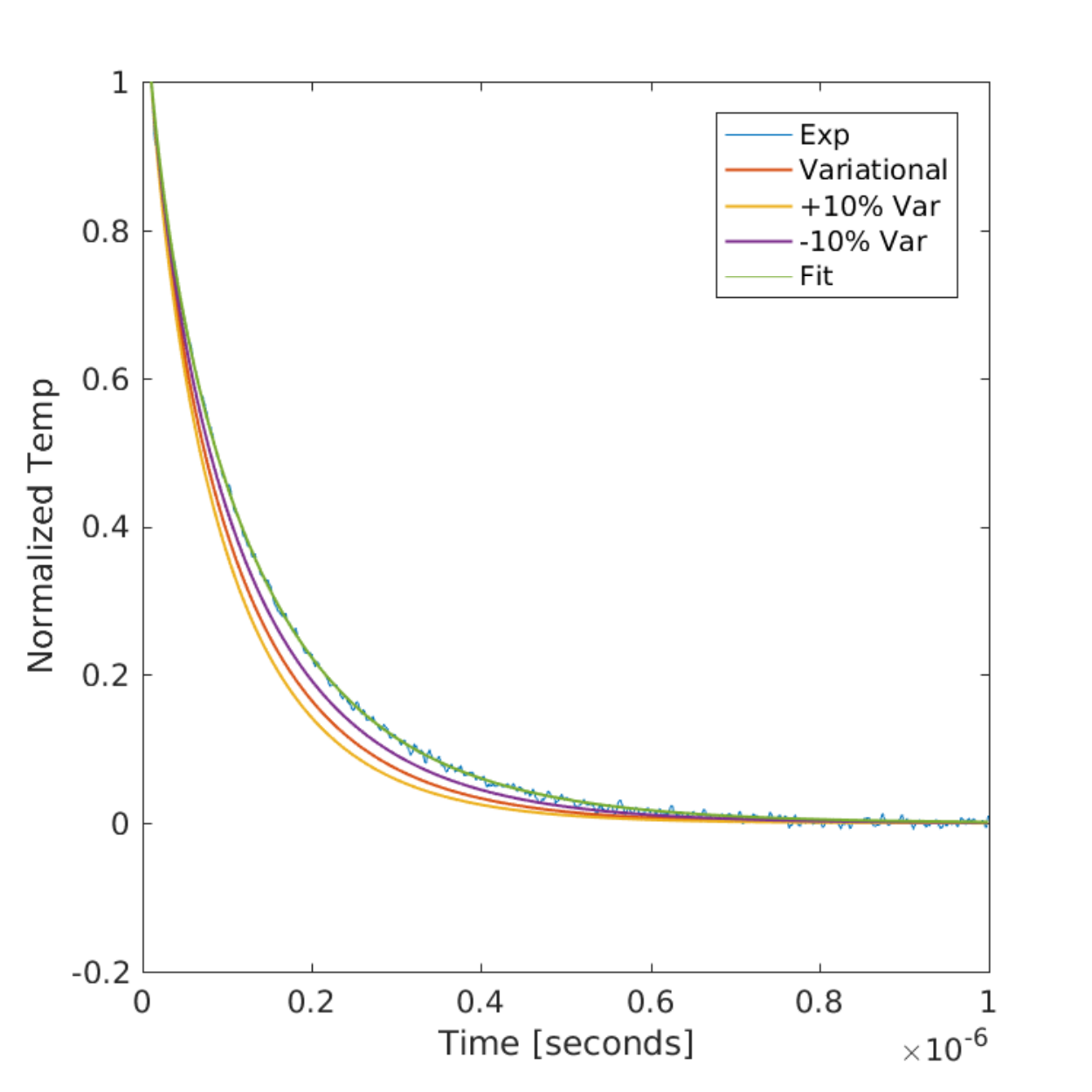}}
\subfloat[4.9 um]{\label{fig:b}\includegraphics[width=0.25\textwidth]{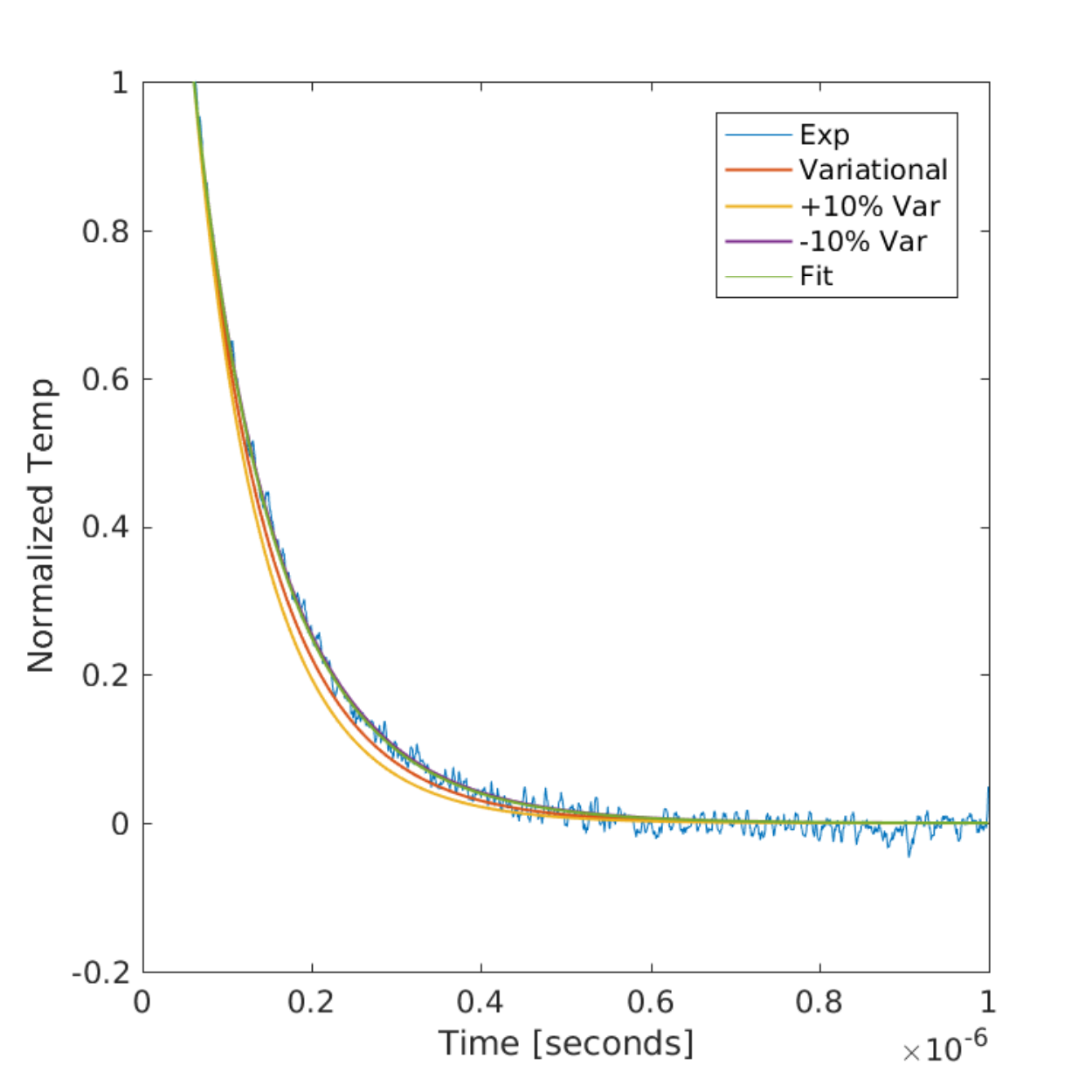}}
\subfloat[4.25 um]{\label{fig:a}\includegraphics[width=0.25\textwidth]{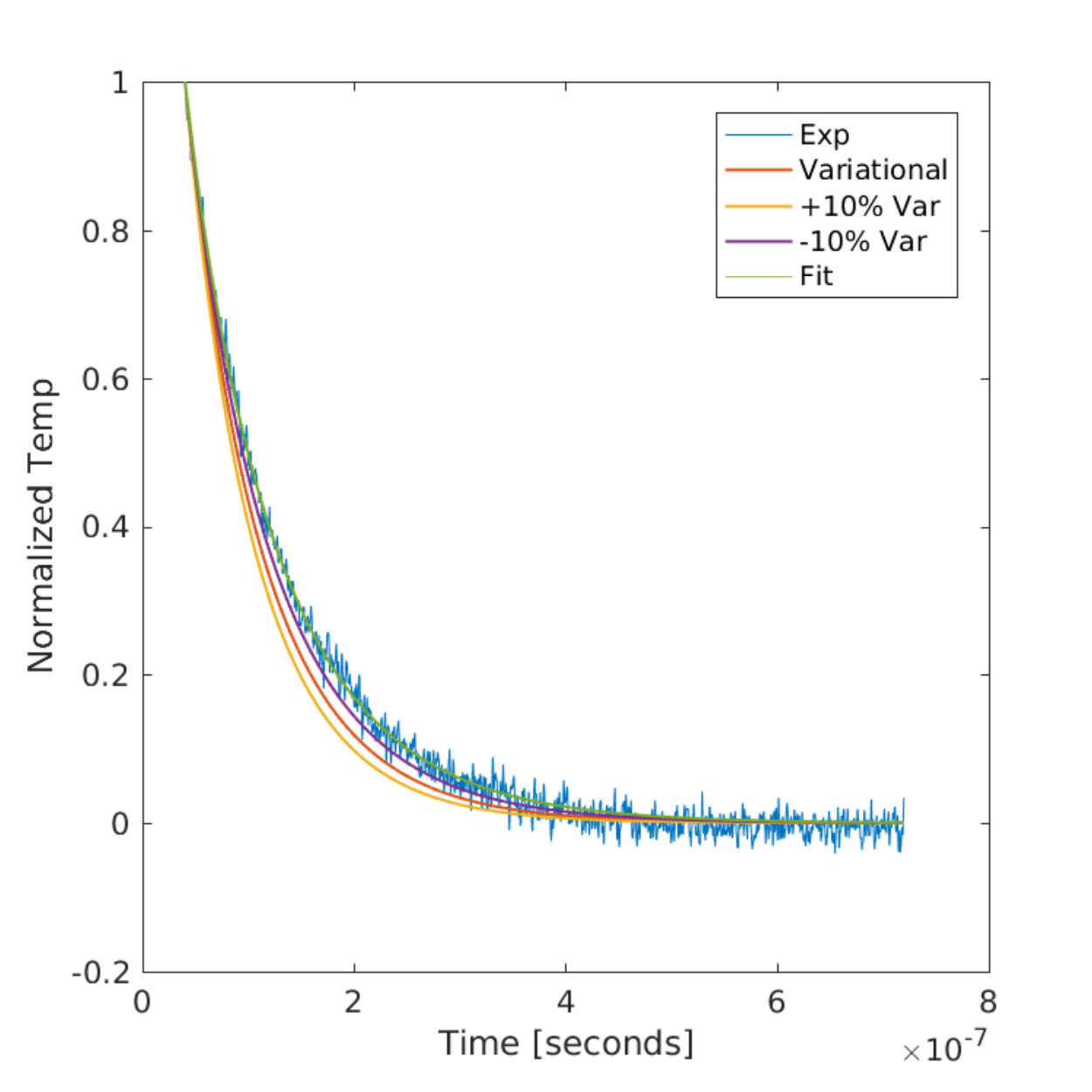}}
\\
\subfloat[3.7 um]{\label{fig:b}\includegraphics[width=0.25\textwidth]{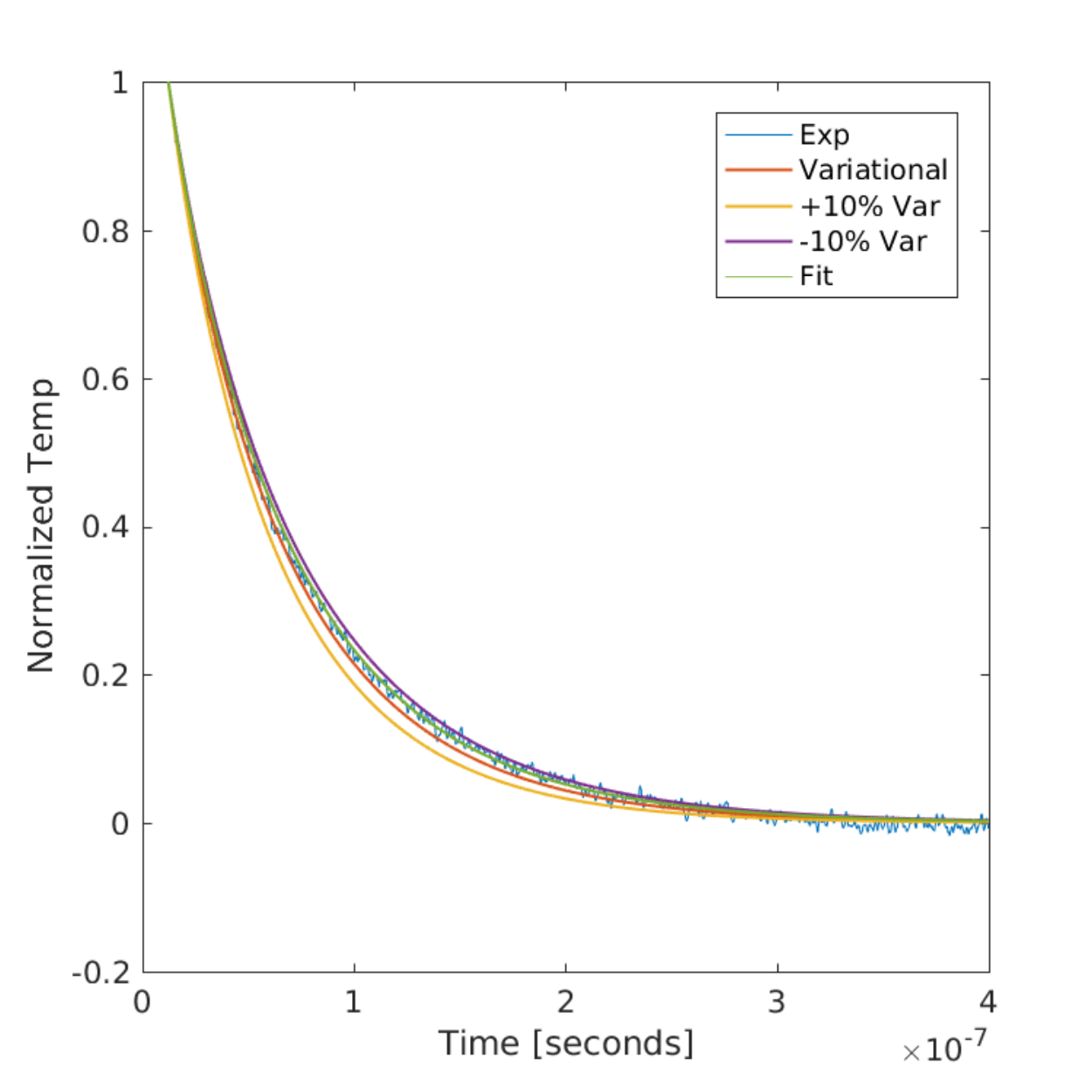}}
\subfloat[2.75 um]{\label{fig:a}\includegraphics[width=0.25\textwidth]{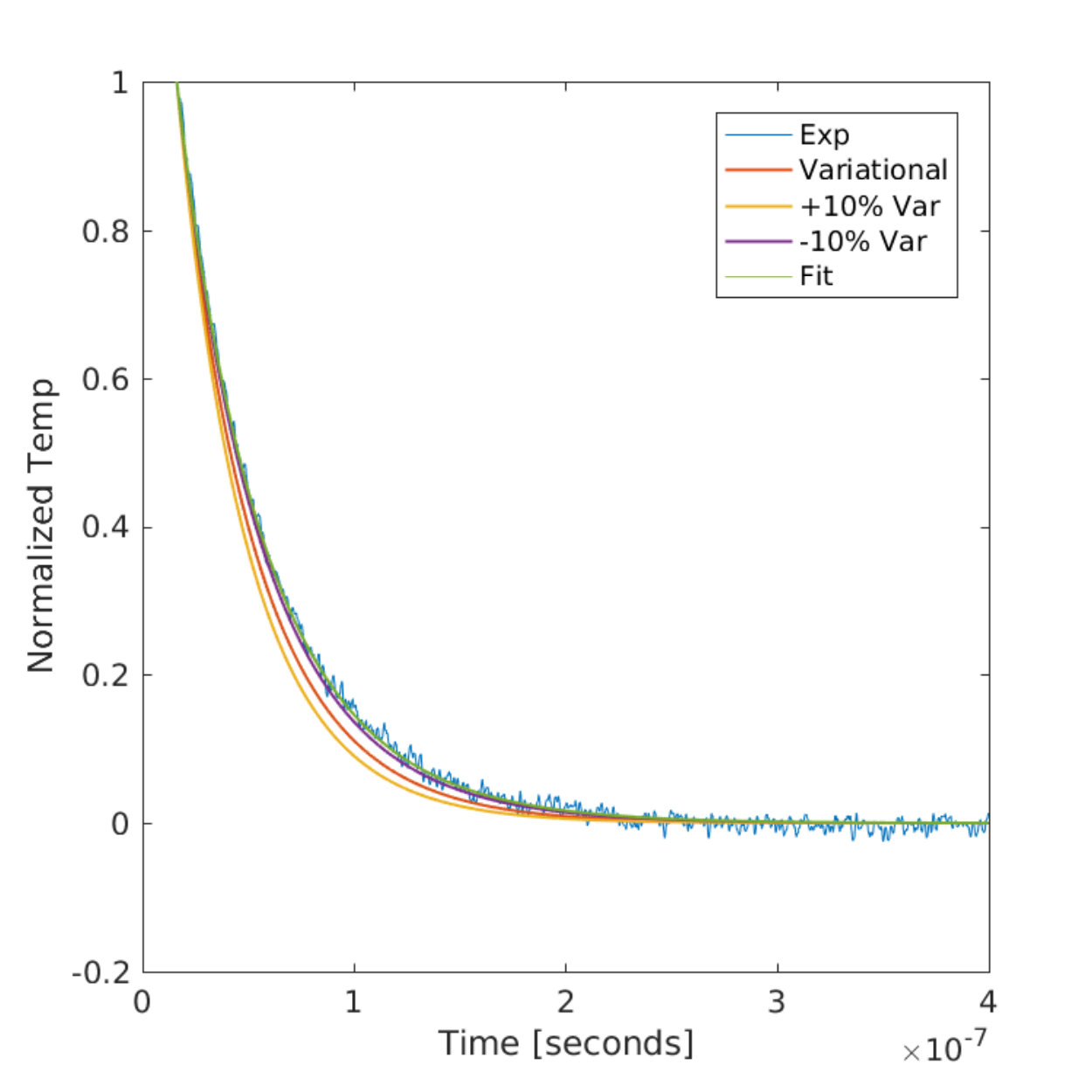}}
\subfloat[2.4 um]{\label{fig:b}\includegraphics[width=0.25\textwidth]{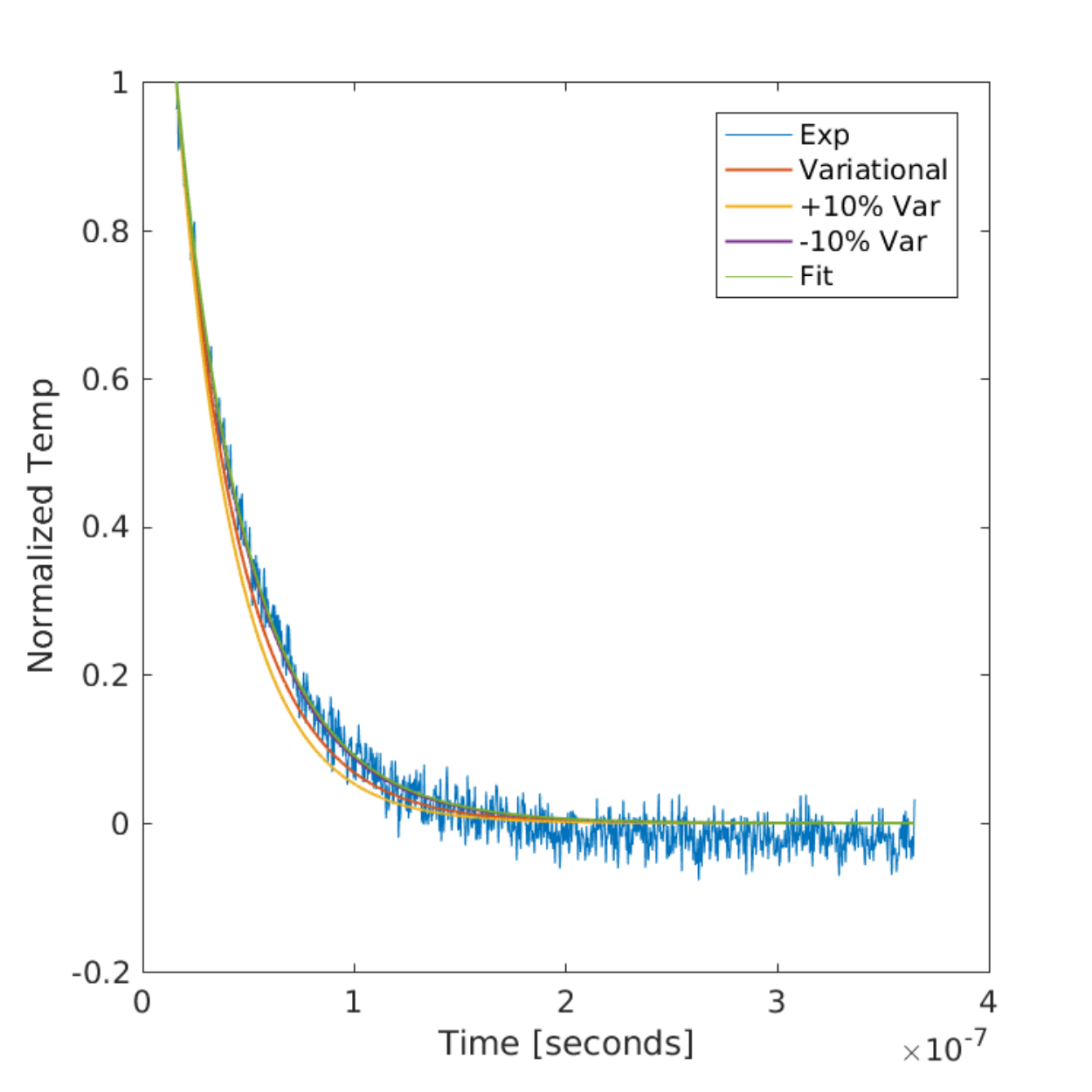}}
\subfloat[2.05 um]{\label{fig:a}\includegraphics[width=0.25\textwidth]{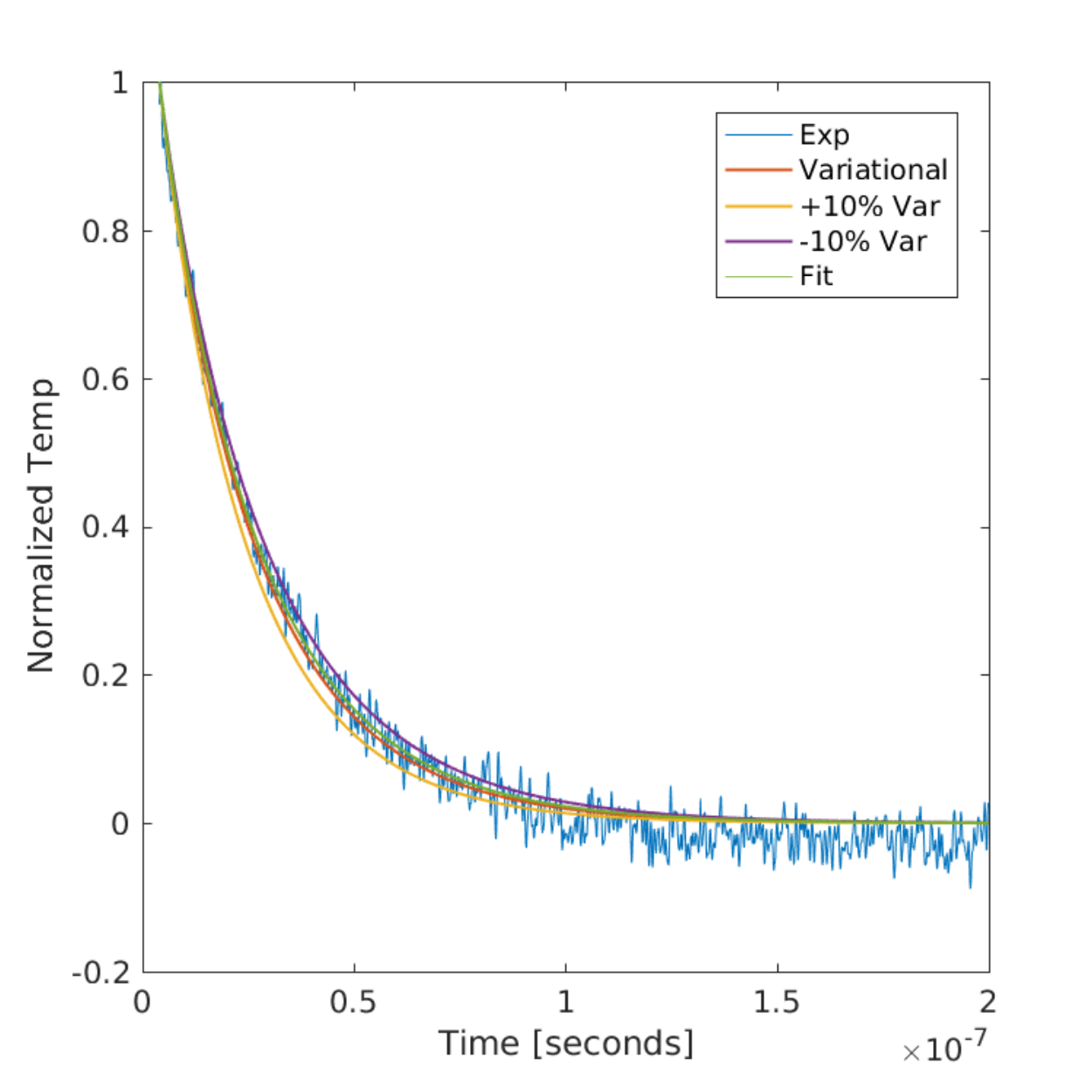}}
\\
\subfloat[1.8 um]{\label{fig:b}\includegraphics[width=0.25\textwidth]{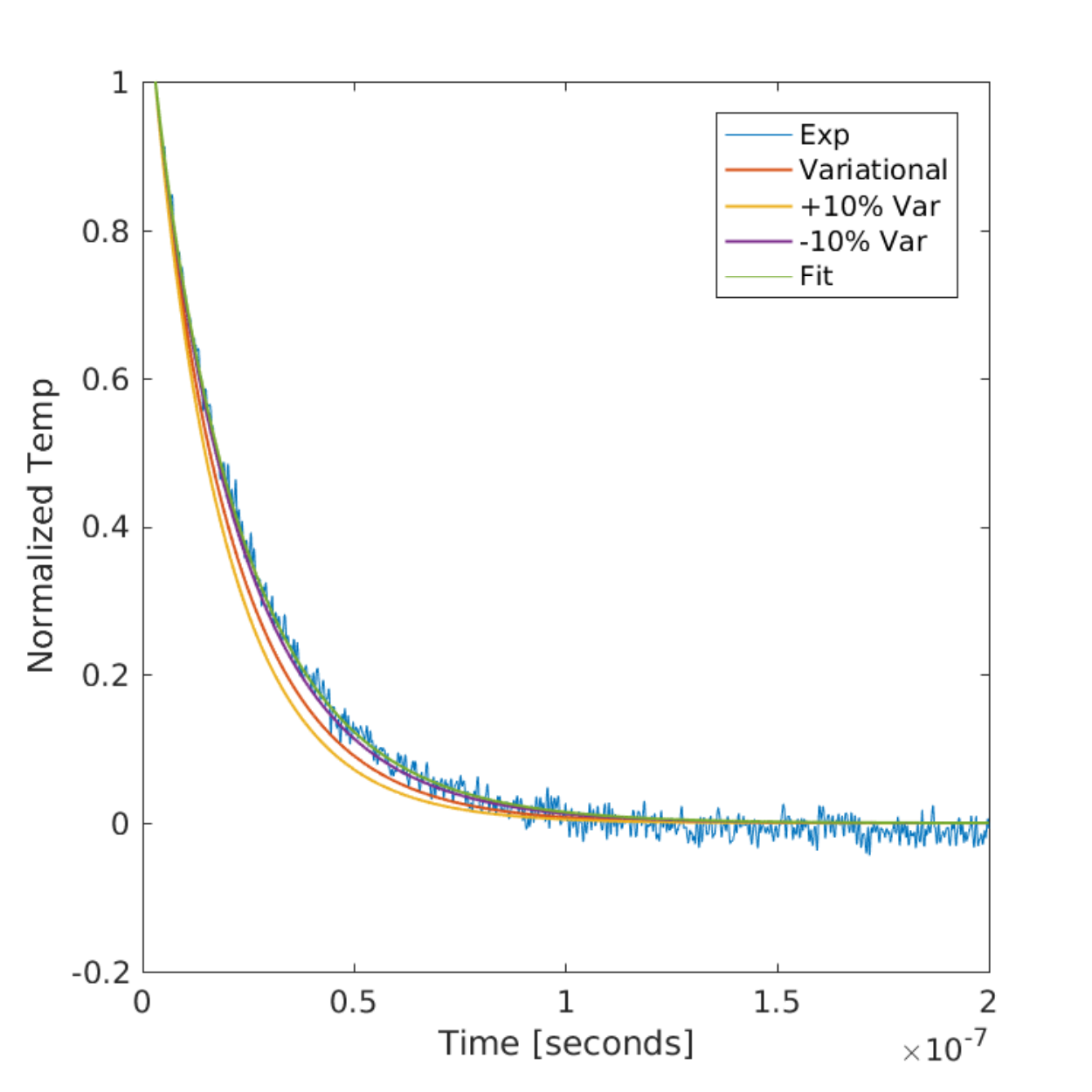}}
\subfloat[1.55 um]{\label{fig:a}\includegraphics[width=0.25\textwidth]{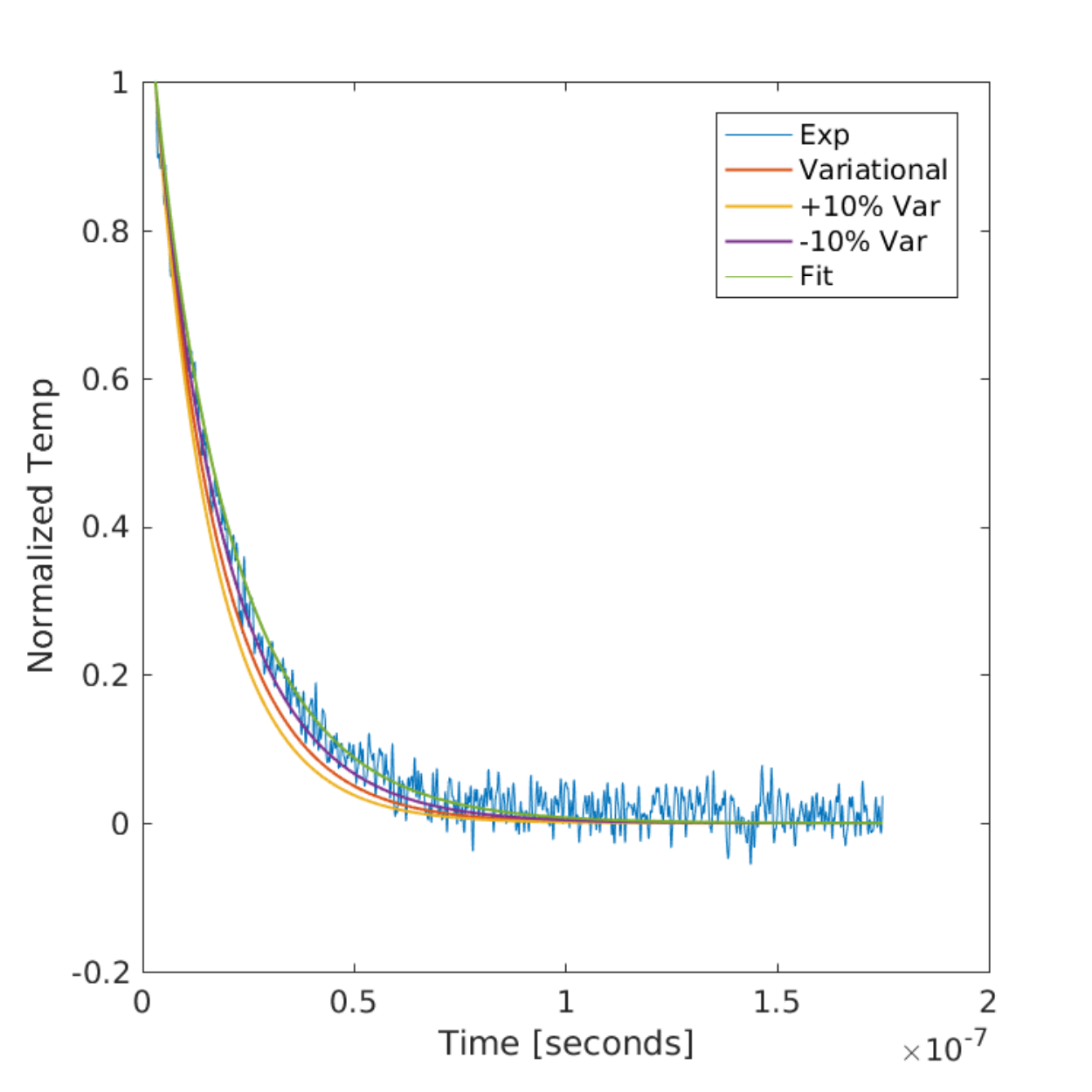}}
\subfloat[1.35 um]{\label{fig:b}\includegraphics[width=0.25\textwidth]{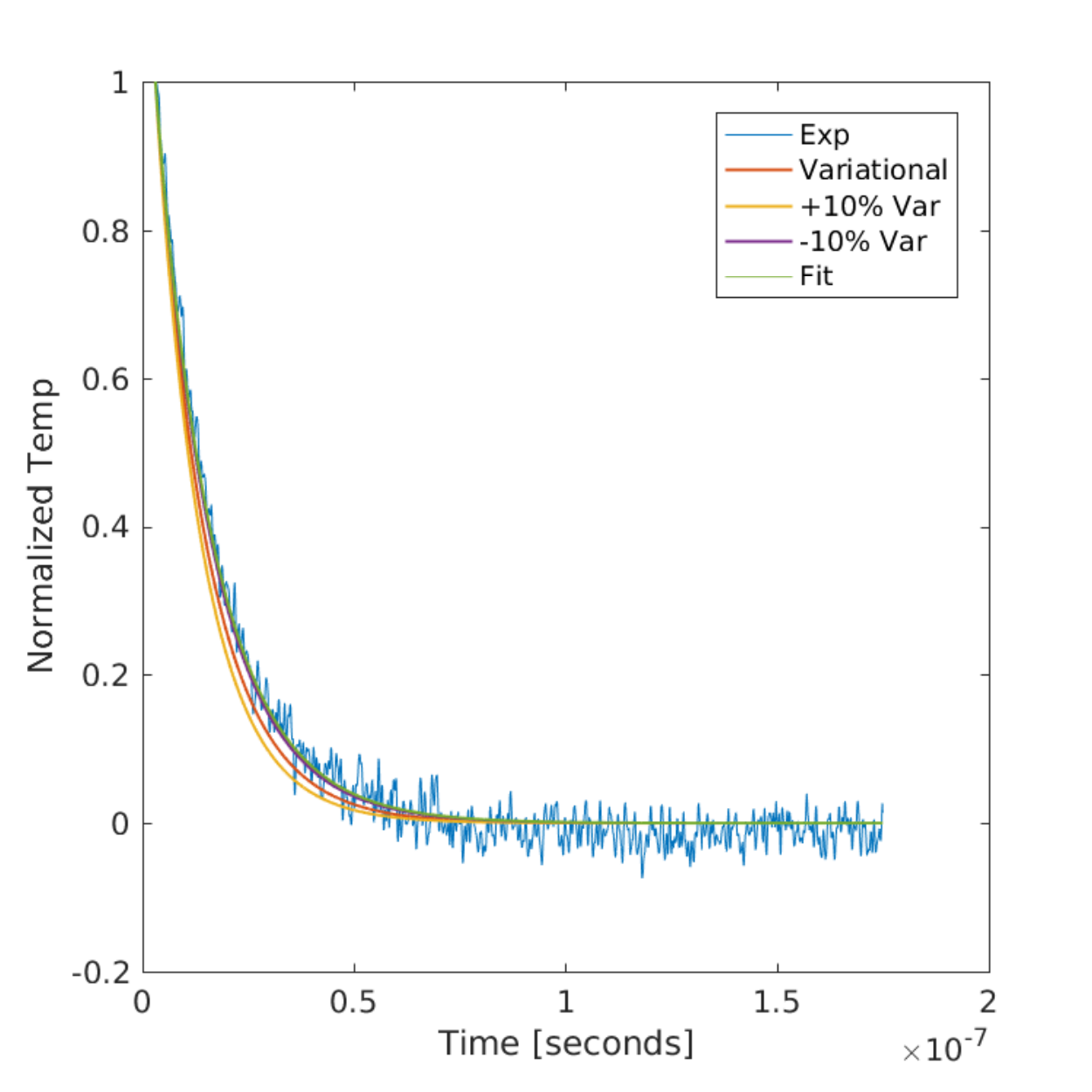}}
\subfloat[1.0 um]{\label{fig:a}\includegraphics[width=0.25\textwidth]{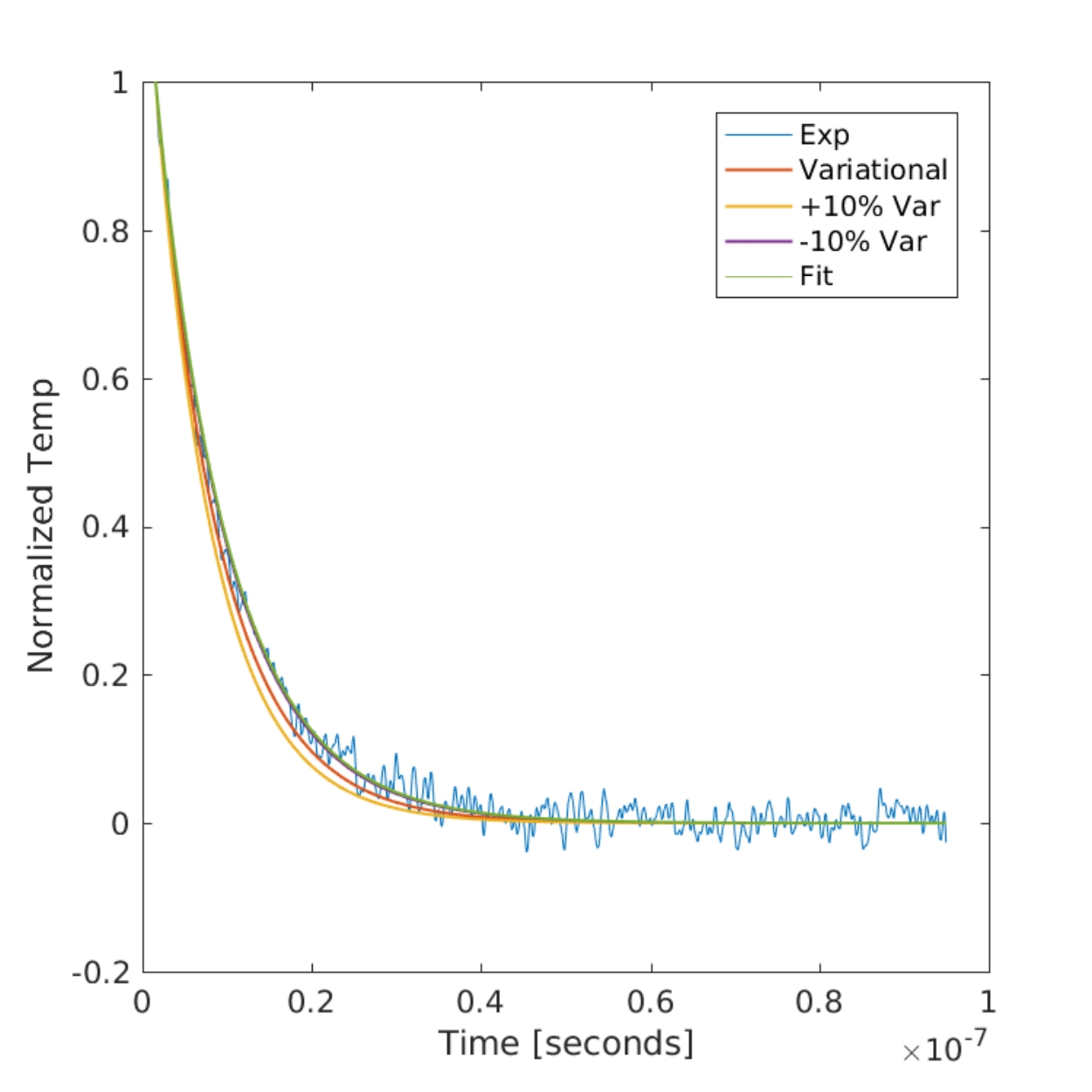}}
\\

\caption{Example TTG results.}
\end{figure}

\newpage
\section{Appendix B}

\begin{figure}[H]
\centering     %%% not \center
\includegraphics[width=0.66\textwidth]{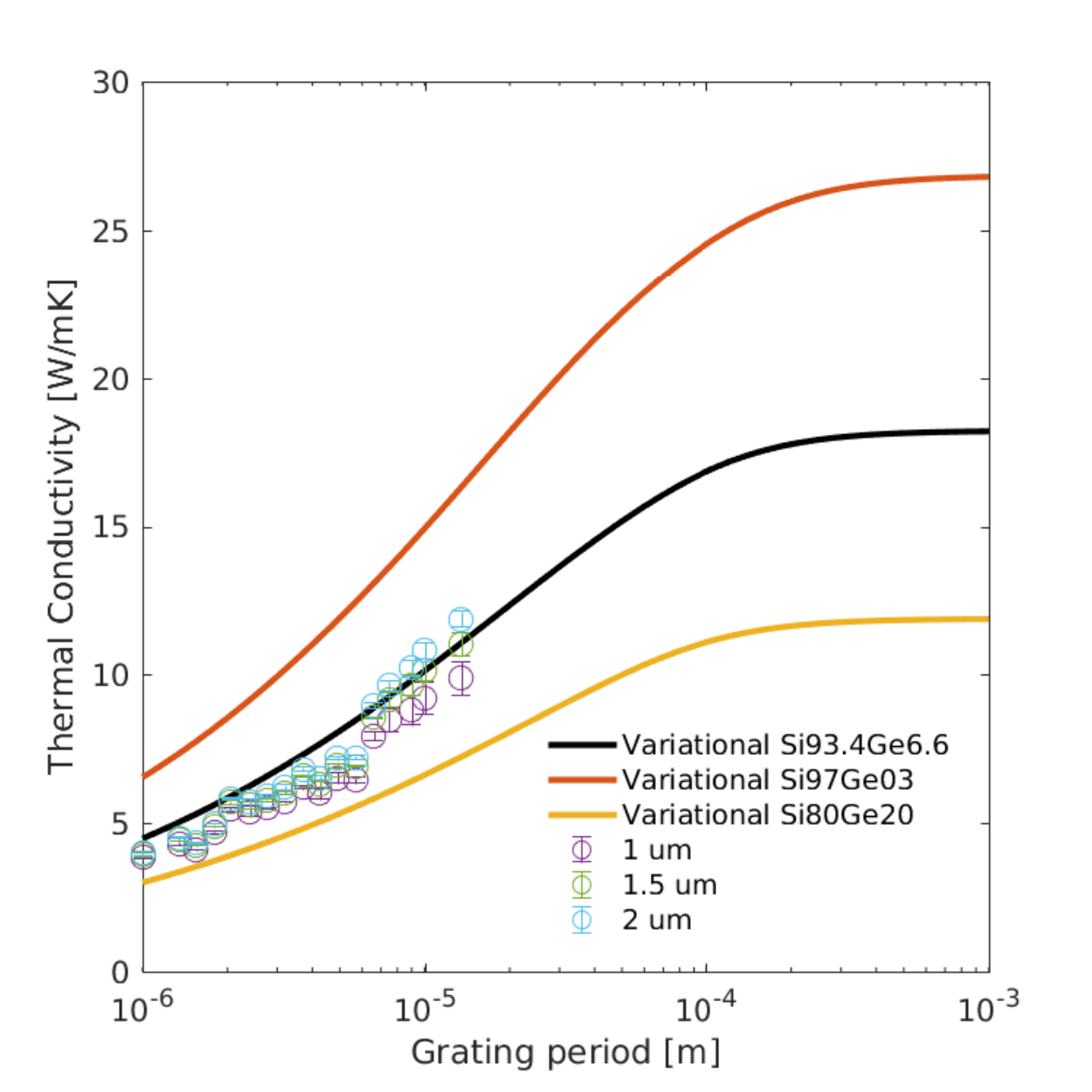}
\caption{Effect of uncertainty in the optical penetration depth on the measurement of effective thermal conductivity with TTG.}
\end{figure}

\newpage
\section{Appendix C}

\begin{figure}[H]
\centering     %%% not \center
\includegraphics[width=0.66\textwidth]{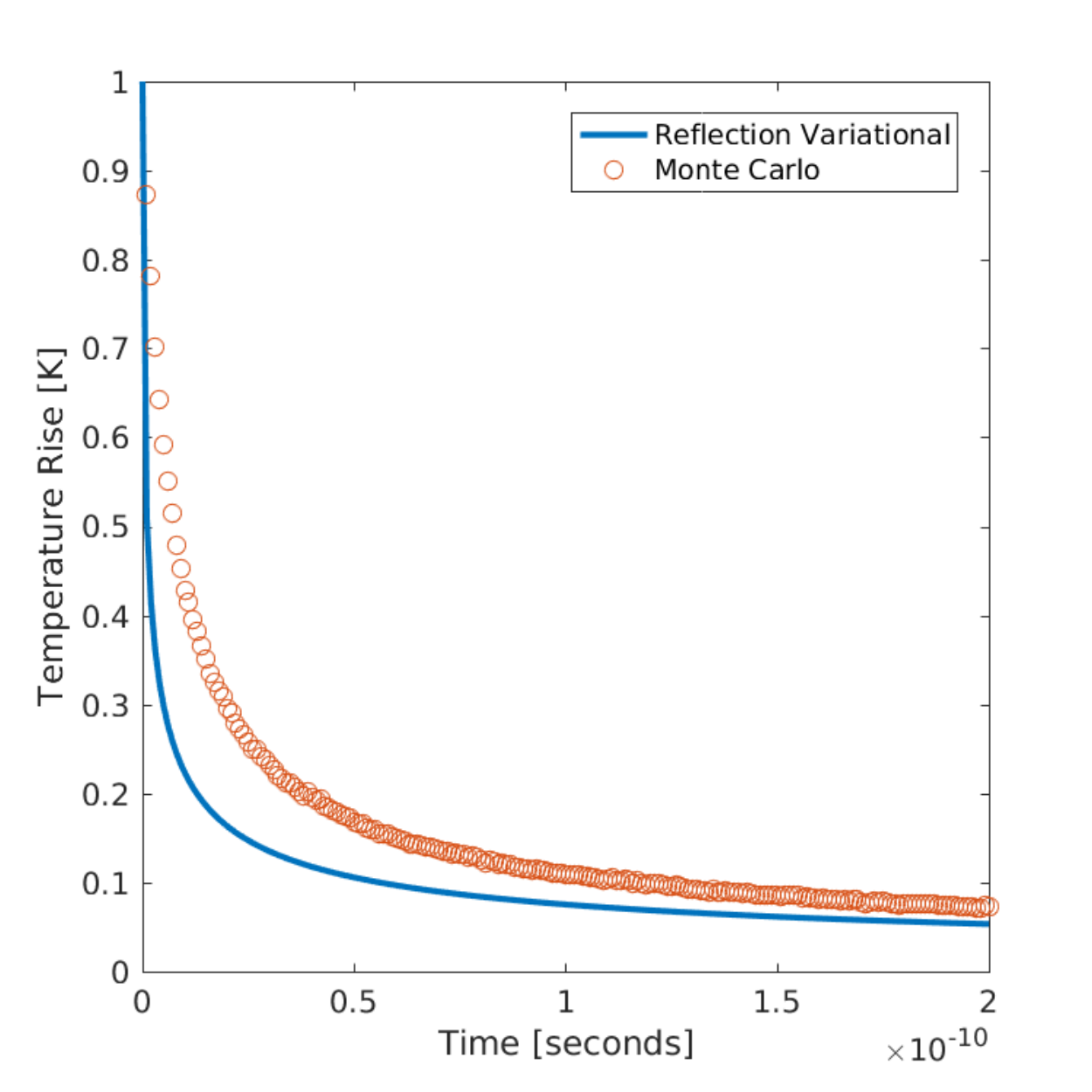}
\caption{Monte Carlo simulation of reflection TTG geometry for Si with grating period of 10 um and optical penetration depth of 10 nm. This example demonstrates the limitations of a modified Fourier treatment, independent of the variational framework. No theory based upon a Fourier model will capture this behavior because a time-dependent thermal conductivity is required to do so and the interpretation of such a quantity can only be phenomenologically understood.}
\end{figure}

\newpage
\section{Appendix E}

\begin{figure}[H]
\centering     %%% not \center
\includegraphics[width=0.66\textwidth]{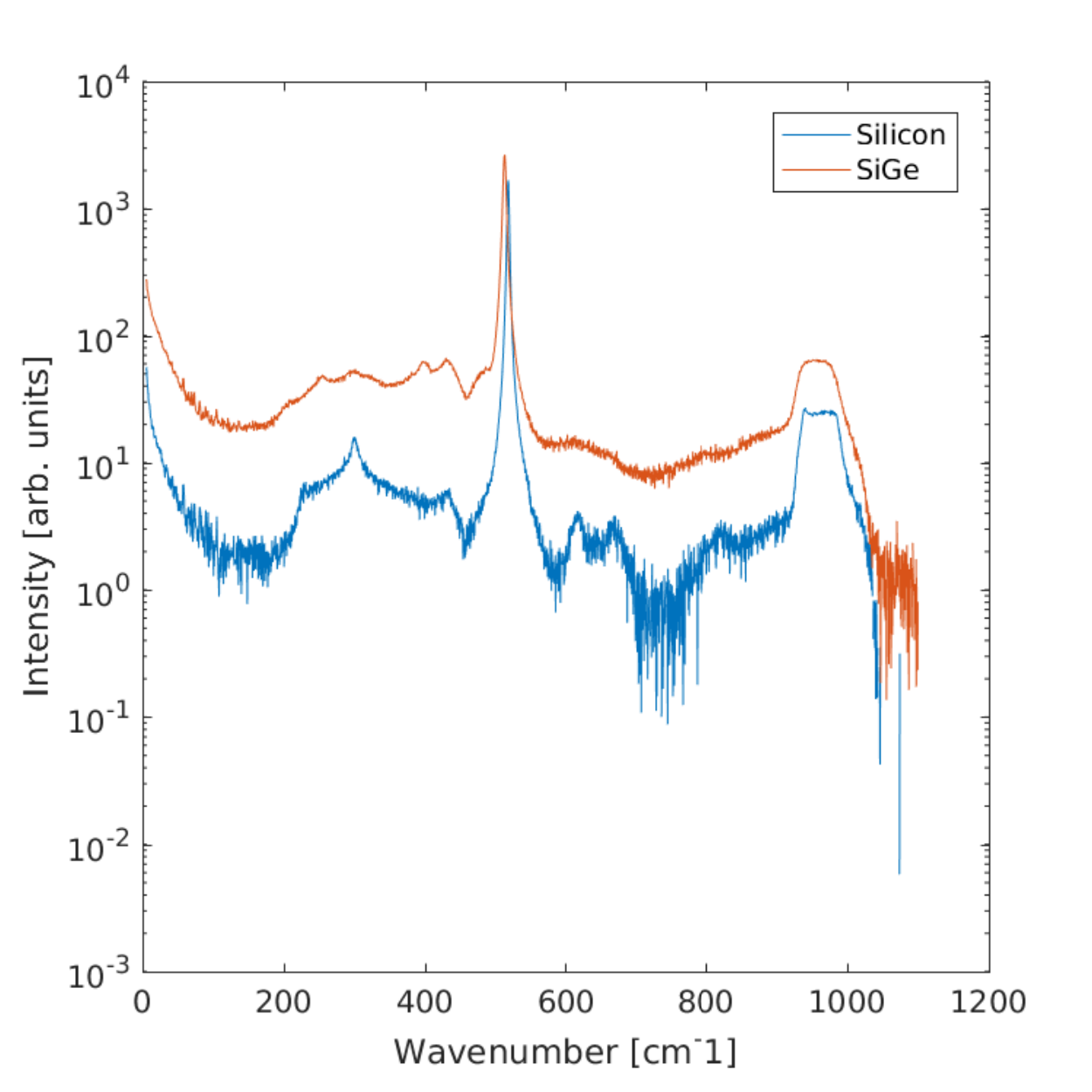}
\caption{Raman spectra for Si and Si$_{93.4}$Ge$_{6.6}$. A dominant peak is found at 520 cm$^{-1}$ for Si and at 513 cm$^{-1}$ for Si$_{93.4}$Ge$_{6.6}$. The DFT-based virtual crystal approximation predicts a single peak at 488 cm$^{-1}$ for Si$_{93.4}$Ge$_{6.6}$.}
\end{figure}
\newpage
\section{Appendix F}

\begin{figure}[H]
\centering     %%% not \center
\includegraphics[width=0.66\textwidth]{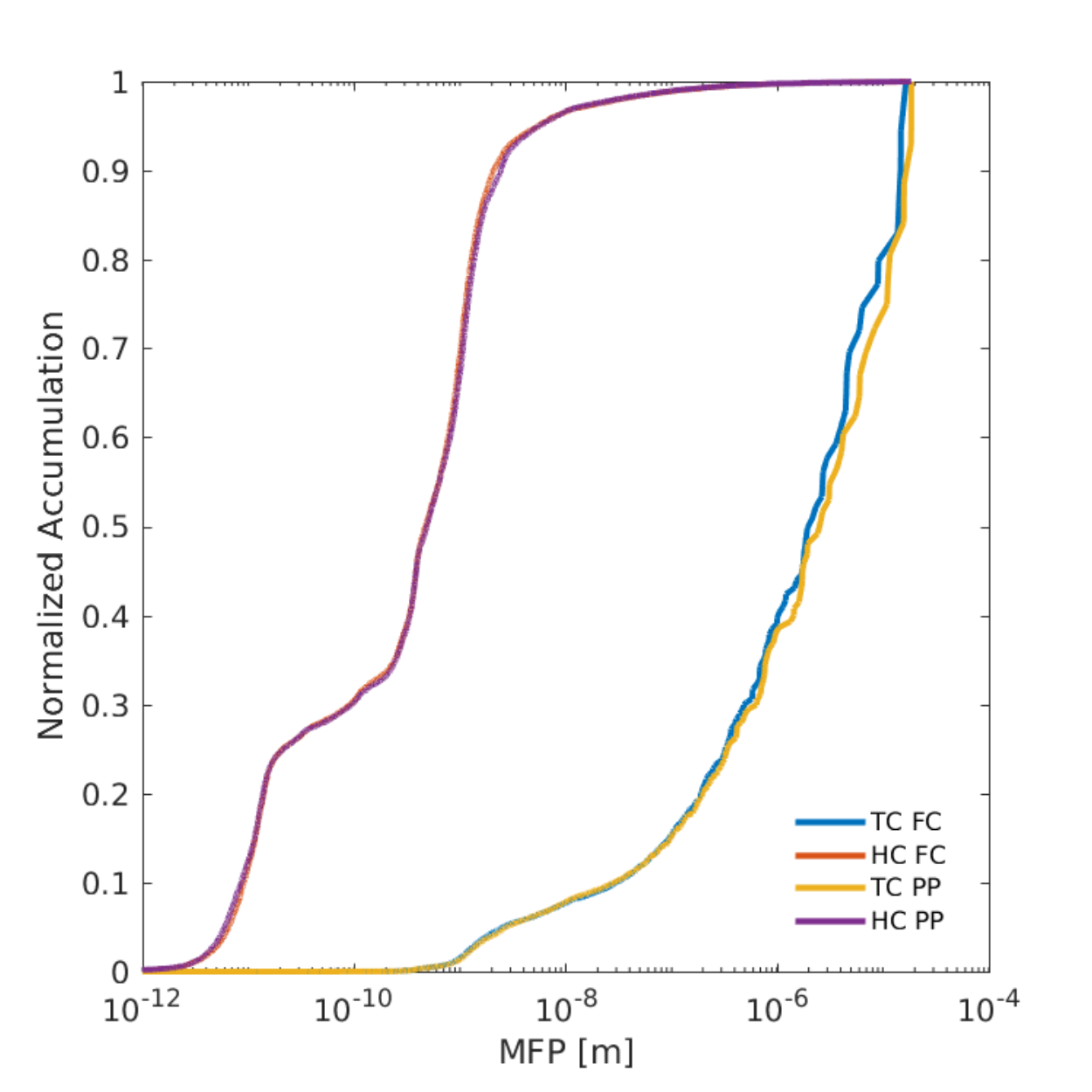}
\caption{DFT comparison between averaging force constants, using the (Si,Ge).pz-bhs.UPF set of pseudopotentials (denoted by FC), and averaging pseudopotentials (denoted by PP) through the virtual.x program, using the (Si,Ge).pz-n-nc.UPF set of pseudopotentials.}
\end{figure}

%\end{document}

\end{document}